\documentclass[11pt,preprint]{aastex}

\newcommand{\ea}{et al.} 
\newcommand{\kms}{\mbox{km\thinspace s$^{-1}$}} 
\newcommand{\cm}{\mbox{$C\!-\!M$}}
\newcommand{\pc}{\mbox{pc}}

\newcommand{\moments}[4]{\mbox{$[{#1}] = {#2}\ ({#3}) \pm {#4}$}} 
\newcommand{\momentsa}[4]{\mbox{$\langle{#1}\rangle = {#2} \pm {#4}$}}
\defcitealias{dru98}{Paper~I}

\begin{document}

\title{The Global Kinematics of the Globular Cluster M92} 
\author{G. A. Drukier}
\affil{ Department of Astronomy, Yale University, New Haven, CT}
\email{drukier@astro.yale.edu}

\author{H. N. Cohn and  P. M. Lugger} 
\affil{Department of Astronomy, Indiana University, Bloomington, IN}
\email{cohn, lugger@astro.indiana.edu}
\author{S. D. Slavin} 
\affil{Dept. of Chemistry and Physics, Purdue University Calumet, Hammond, IN}
\email{slavin@calumet.purdue.edu}
\author{R. C. Berrington}
\affil{Dept. of Physics and Astronomy, University of Wyoming, Laramie, WY}
\email{rberring@uwyo.edu}
\and
\author{B. W. Murphy} 
\affil{Department of Physics and Astronomy, Butler University, Indianapolis, IN}
\email{bmurphy@butler.edu}

\begin{abstract} 

We report the determination of high-accuracy radial velocities for
299 members of the globular cluster M92 using the Hydra
multi-object spectrograph on the WIYN telescope.  We have
concentrated on stars outside of the  central region of the
cluster, located up to 14\farcm 4 from the cluster center.  Candidate
members were selected for spectroscopy based on a photometric
metallicity index determined from 3-band Washington photometry, also
obtained with the WIYN telescope. The median error in the velocities
is 0.35 \kms. We find the heliocentric radial velocity of the cluster
to be $-121.2 \pm 0.3\ \kms$. 

We have used an improved Bayesian analysis to determine the velocity
dispersion profile of M92\@.  The most probable profile is a cored
power-law with a scale radius of 2\arcmin, velocity dispersion at
1\arcmin\ of 6.3 \kms\ and outer power-law with slope $-0.6$.  We have
also reanalyzed the M15 radial velocities of \citet{dru98} and find
that a pure power-law with a 1\arcmin\
velocity dispersion of 8 \kms\ and slope $-0.5$, and the combination
of a power-law with slope $-0.4$ and scale of 7.5 \kms\  inside 9\arcmin\
and a dispersion of 4 \kms\ outside, are equally likely. In both
clusters there is  evidence that the samples include
escaping stars. We present results from a GRAPE-based N-body
simulation of an isolated cluster that demonstrates this effect.  We
suggest additional tests to determine the relative importance of tidal
heating and stellar ejection for establishing the velocity field in
globular cluster halos.

\end{abstract}


\section{Introduction} 

In globular clusters, the interplay between two-body relaxation and
external tidal stresses is most obvious in their outer parts. There
has been considerable recent interest in the evolution and eventual
dissolution of clusters in the Galactic tidal field
\citep*[e.g.][]{gne99,com99,takp00,de04}. To investigate these
issues, we have been carrying out a program to determine the global
velocity fields of globular clusters using the Hydra fiber-fed,
multi-object spectrograph on the WIYN telescope\footnote{The WIYN
Observatory is a joint facility of the University of Wisconsin,
Indiana University, Yale University, and the National Optical
Astronomy Observatories.}.   With
its 1\arcdeg\ diameter field and echelle grating, Hydra is well suited
for determining high-accuracy radial velocities of stars out to the
tidal radii of clusters.  Our approach is an important complement to
Fabry-Perot imaging \citep[e.g.][]{geb94,geb00}, which can only be
used efficiently in the central $1-2\arcmin$. 

In a previous paper \citep[hereafter Paper~I]{dru98}
we
reported a new global velocity-dispersion profile for the prototypical
collapsed-core cluster M15.  We found a clear indication of a flattening
and possible rise of the profile in the outer part of the cluster.  In
contrast, our anisotropic Fokker-Planck simulations of isolated clusters
show a smoothly declining velocity-dispersion profile \citep{dru99}.  We
interpreted our observations of M15 as evidence for heating of the
cluster halo by the Galactic tidal field.  \citet{joh99} subsequently
showed that a Galactic satellite system undergoing tidal heating will
show a break in the slope of the velocity dispersion profile at the
radius at which unbound stars begin to dominate.  However, this leaves
open the question of whether two-body relaxation in an isolated cluster
could produce a similar effect through \emph{strong} scattering of stars
from the central region into unbound orbits.  Neither our Fokker-Planck
simulations nor the self-consistent field simulations of \citet{joh99}
include strong scattering.  We address this question in the present
study through direct N-body integrations, which include all scattering
mechanisms. 

We chose M92 (NGC~6341) for the second cluster in our survey, as a
``normal'' comparison to the collapsed-core cluster M15\@.  The
properties of these two clusters are compared in
Table~\ref{T:M92-M15}.  Unlike M15, M92 has a well-resolved core.
While its central luminosity density is in the upper third of the
distribution for all Galactic globular clusters \citep{har96}, it is
nevertheless at least an order of magnitude lower than that of M15\@.
Both clusters are brighter than the median of the absolute magnitude
distribution; M92 is in the upper 30\%, while M15 is in the upper
decile.  Both are among the very lowest metallicity clusters.  Both
are located approximately 10~kpc from the Galactic center and nearly
5~kpc from the Galactic plane.  Both are among the oldest group of
clusters \citep{cha96}.

The radial velocity of M92 ($-121.2\pm 0.3$~\kms) and its very low
metallicity ([Fe/H]$=-2.29$) clearly distinguish its stars from the
bulk of the field stars in the same region of the sky.  Nevertheless,
selecting likely cluster members for spectroscopy becomes challenging
in the outer halo where the vast majority of stars are nonmembers.  As
we pointed out in \citetalias{dru98}, selecting stars based on their
position in a color-magnitude diagram (CMD) alone is not an efficient
means of finding potential members.  As we discuss in the next
section, we turned to a metallicity-sensitive color-color diagram in
the Washington system \citep{can76} to aid our candidate selection.

There have been a number of previous velocity studies of M92. The most
extensive is that of \citet*{lup85}, who observed 49 stars in M92
using the 5~m Hale telescope, but, unfortunately, the individual
velocities have not been published. \citet{lup85} find little sign of
anisotropy and only a weak ($2\sigma$ at best) rotational signal. The
dispersion profile itself is fairly flat,
but the statistics in the outer part of the cluster are particularly
poor.  \citet{bee90} observed only 7 stars and the uncertainty in the
velocities is 7~\kms.  While the present study was underway,
\citet{sod99} reported high-precision measurements for 35 stars in
M92, based on Hydra spectroscopy with the KPNO~4m and the WIYN
telescope in the region about the Na~D line.  Thus, while some
previous radial velocity information is available for M92, it is not
sufficient for a determination of the global velocity dispersion
profile, particularly in the cluster halo.  Good quality proper
motions for M92 stars are available for of order 700 stars located out to
19\arcmin\ from the cluster center \citep*{ree92,tuc96}.

As in \citetalias{dru98}, we carried out our observations on the 3.5~m
WIYN telescope.  We used the Hydra
fiber-fed, multi-object spectrograph to obtain high-resolution
spectra.  This has resulted in a homogeneous set of over 300 M92
stars with new, high-accuracy, velocity measurements.

In the absence of a full dynamical study, data sets such as these
are customarily analyzed by binning the stars in radius and
computing the dispersion in their velocities. In this way, an
approximation to the dispersion profile is constructed. In this
paper, we use the methods of Bayesian statistics to move beyond this
 in three ways. First, besides the traditional bins, we also examine
 several  continuous dispersion profiles as described in
\S\ref{models}. Second, our methods allow us to assess which of the
several candidate profiles is best supported by the data. Third, by
looking at the posterior probability distributions for the
parameters we can determine the extent to which they deviate from
Gaussian distributions. 

We discuss our observations in the following section.  It particular,
we review our new photometric selection technique using the Washington
system.  In \S\ref{kinematics} we use the Bayesian techniques
described in \S\ref{analysis} to analyze the velocity distributions of
both the new M92 data and the M15 data from \citetalias{dru98}.  We
then do a similar analysis of an 8k N-body model to investigate the
effects of unbound stars on the dispersion analysis.  We conclude with
a summary of the main results of this paper and a suggestion of some
directions for further progress in investigating the questions
raised by these data sets.

\section{Observations} 
\subsection{Photometry and Candidate Selection} 

In order to observe with Hydra, it is necessary to prepare a list of
target stars with accurate positions. For this project, the targets
are those stars found on the giant branch which are likely to be
cluster members.  Our candidate list comes from three sources. For our
1996 May run and the first night of the 1996 June run, our candidates
all came from the proper motion list of \citet{ree92}.  We selected
stars appearing on the $V$ vs.\ \bv\ CMD, but did not select on the
basis of the assigned membership probabilities. For the bulk of the
1996 June run, our stars were selected from a list of stars found on
KPNO Burrell Schmidt CCD images kindly supplied by H. Morrison. The
images had been taken in the Washington system, which we discuss
further below.  Owing to problems with obtaining sufficiently accurate
photometry from these images, we selected stars only on the basis of
their CMD position, rather than using the color-color plot which
provides metallicity information.  We observed 444 stars in 1996 and
obtained useful velocities for 427 of these. Only 192 stars (45\%)
proved to be members.

As we discussed in \citetalias{dru98}, selecting stars on the basis of
a CMD alone is not the most efficient method of finding likely
members.  Depending on the distance and Galactic latitude of the
cluster, contamination by Galactic dwarfs with the colors of the
cluster giant branch can be significant. This is a particular problem
in the outer part of a globular cluster where the members are sparse
on the sky and are outnumbered by non-members even at high Galactic
latitudes. A better method is to take advantage of the metallicity
difference between a globular cluster and the Galactic field.

The Washington system \citep{can76} is designed to determine stellar
metallicities along the giant branch. It consists of four broad-band
filters, $C$, $M$, $T_1$, and $T_2$. The last is equivalent to
Kron-Cousins $I$.  The $T_1\,-\,T_2$ color was originally intended to
be the temperature index, with \cm\ giving the metallicity. The $C$
band was selected to occupy the region around 400 nm occupied by the Ca
H and K doublet, the G band, and plentiful CN bands in these cool
stars. The $M$ band is in a spectral region centered near 510 nm which
is less affected by metallicity.  \citet*{gei91} have shown that the
optimal diagram for metallicity determination is the \cm\ vs.\
$M\!-\!T_2$ color-color diagram.  This gives us the maximum
metallicity sensitivity and only requires observation in three
broad-band colors.

In 1997 May, we observed M92 using the imager on the WIYN telescope.
We observed a 40-field mosaic covering a 15\arcmin\ radius centered
on the cluster, with a few fields extending out to 20\arcmin. We
observed each field with the $C$, $M$, and $T_2$ filters.  We also
observed using the DDO~51 filter as suggested by \citet{gei84}. This
should have allowed us to separate dwarfs from giants on the basis of
the luminosity effects in the Mg~b triplet. The DDO~51 photometry
lacked sufficient precision to permit this, possibly due error
propagation in transferring the zero-point across the large mosaic. The
DDO~51 frames were used for the astrometry described below.

Each frame was debiased and flat-fielded in the usual manner, and
photometry was done on each using DAOPHOT~II and ALLSTAR
\citep{ste87,ste88}.  Since the imaging was not done under photometric
conditions, all of the photometry was put on a common zero-point by
matching stars in overlapping regions between frames.  In a
subsequent, photometric, run in 1997 October, we observed one field in
M92 as well as a region in  standard field SA114 containing three standard stars
from \citet{gei96}.  This allowed us to put our $C$, $M$, and $T_2$
observations onto the standard system. Having only the one standard
image, we used the first order extinction coefficients from
\citet{har79}.  Our selection of likely cluster members, however, was
based on the uncalibrated color-color diagram.

\begin{figure}
\plotone{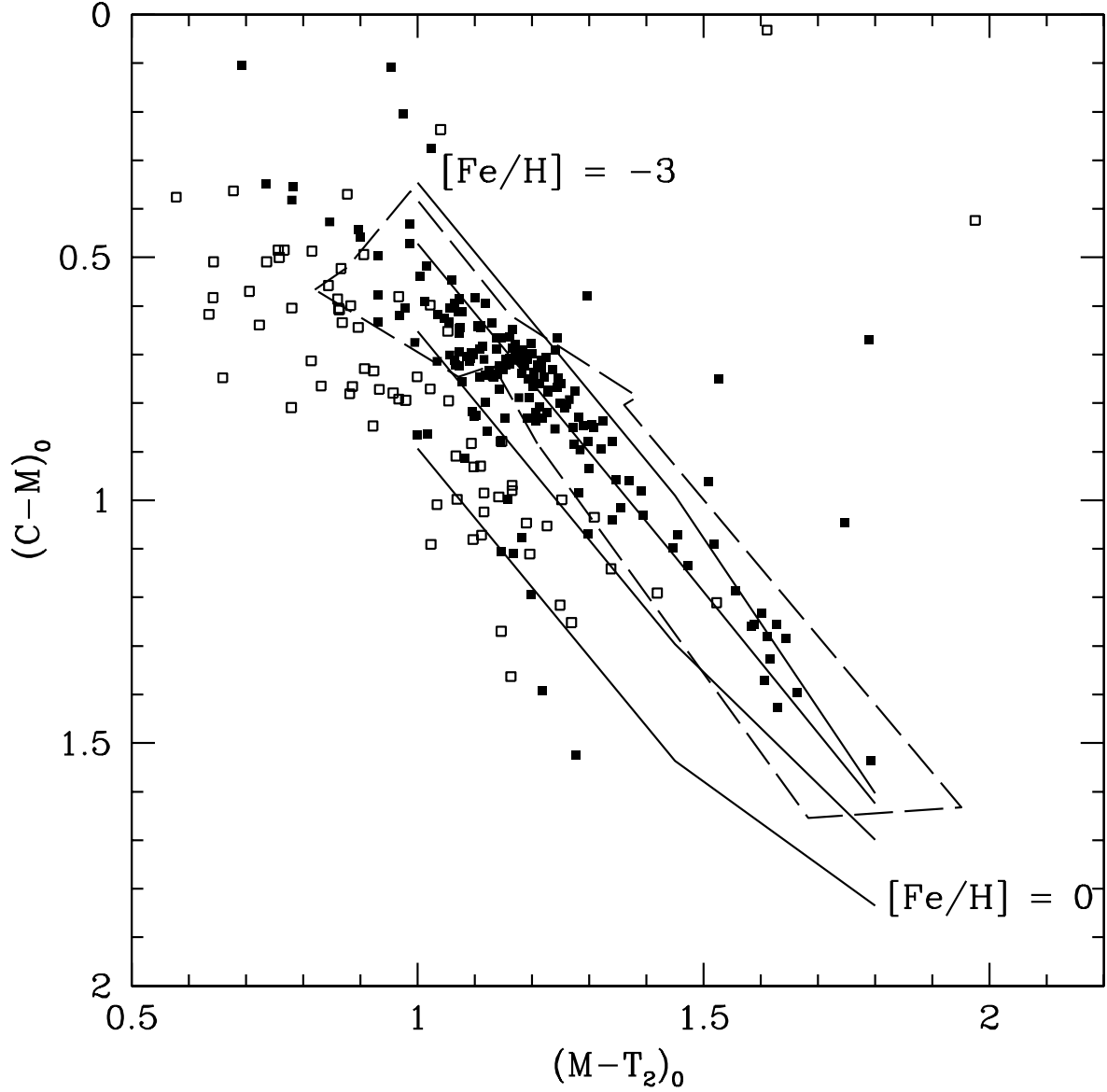} 
\figcaption[f1.eps]{Washington system \cm\ vs.\
$M\!-\!T_2$ color-color diagram for the stars observed in 1996. Only
stars with the higher-precision, WIYN photometry are shown. The filled
symbols are the stars which were subsequently determined to be members
of M92 and the open symbols are non-members. The solid lines are locii of common
metallicity from Geisler (1991). Based on this diagram, we selected
the region within the dashed lines for our 1997 observations.
\label{F:ccd96}}
\end{figure}


\begin{figure}
\plotone{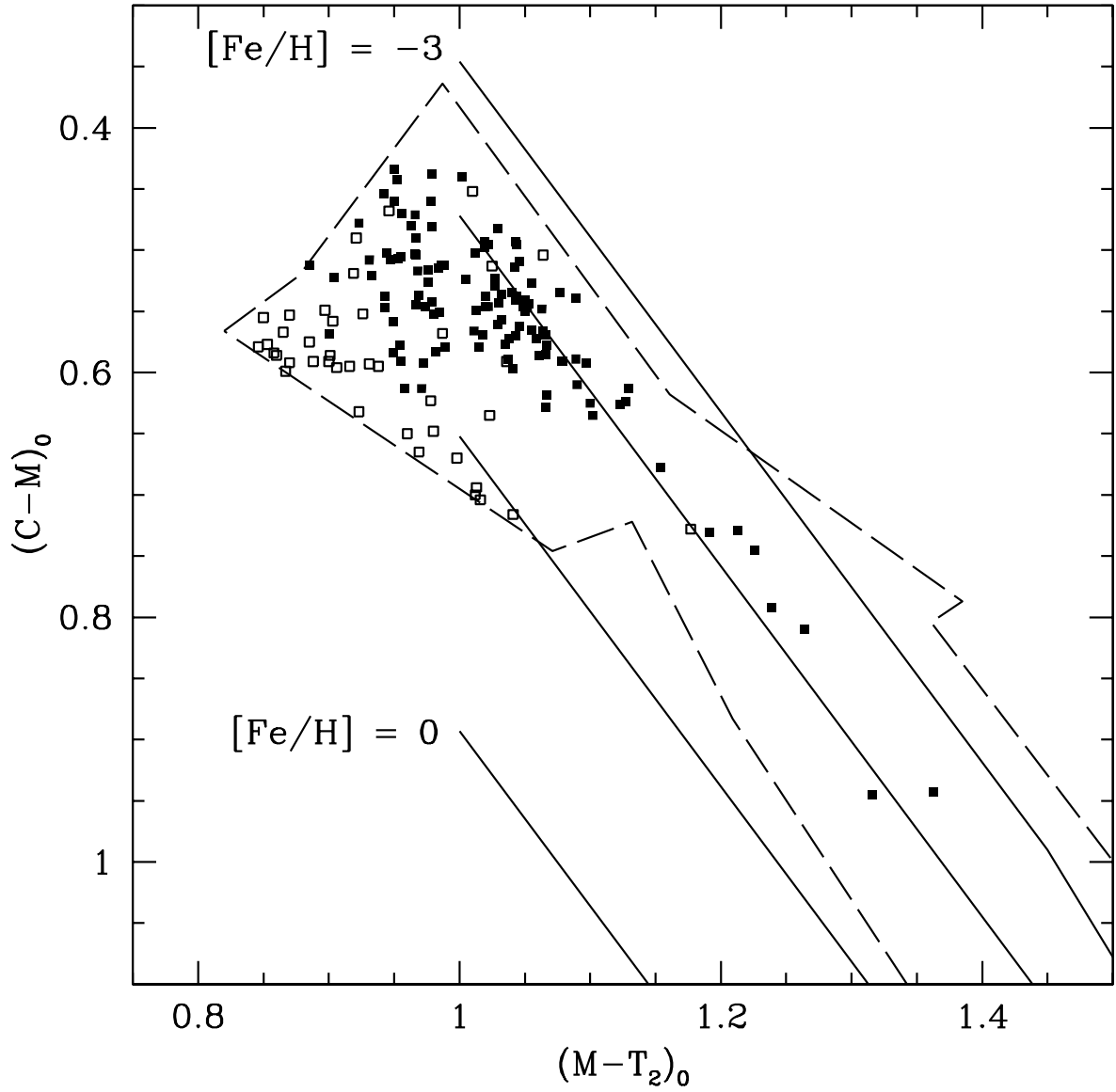} 
\figcaption[f2.eps]{As Figure~\protect{\ref{F:ccd96}}
for the stars observed in 1997. This stars in this sample were selected
only on the basis of their positions in this diagram. 
\label{F:ccd97}}
\end{figure}

 \begin{figure}
\plotone{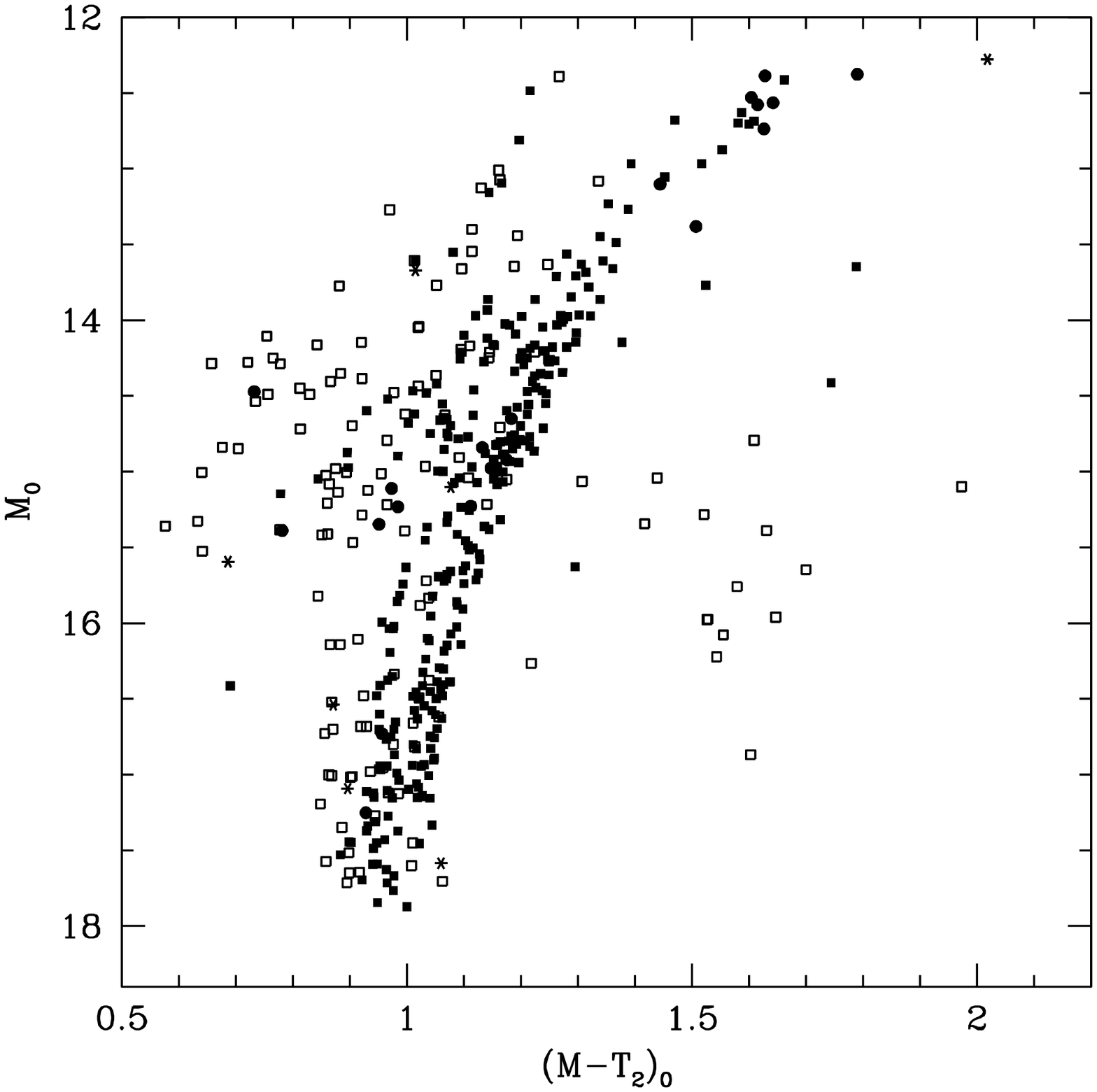} 
\figcaption[f3.eps]{Washington system color-magnitude
diagram for all stars with measured velocities. The filled symbols are
confirmed members, with the  circles for the velocity variables
and the  squares non-variables. The open squares are non-members and
the stars doubtful members. Most of the stars brighter than $M=15.5$
were observed in 1996; those fainter are from the color-selected
1997 sample. 
\label{F:cmd}}
\end{figure}

Our original region, selected to include the giant branch, was $M<18$,
$\cm>0.2$ and $M\!-\!T_2>0.5$. The stars observed in 1996 with
photometry from our WIYN mosaic are shown in Fig.~\ref{F:ccd96} in the
calibrated \cm\ vs. $M\!-\!T_2$ color-color diagram. We have adopted
$E(\bv)=0.02$ \citep{har96} for M92 and have taken the reddening
ratios from \citet{har79}.  These stars were selected on the basis of
their position in the color-magnitude diagram. The open symbols are
the non-members and the filled symbols are the
spectroscopically-determined members. We have also included the locii
of the ${\rm [Fe/H]}=0$ through ${\rm [Fe/H]}=-3$ stars, in 1 dex
steps in ${\rm [Fe/H]}$,
from Fig.~7 of
\citet{gei91}.  The separation by metallicity is quite clear.  We
based our final selection of stars for the 1997 observing season on
their location in this diagram.  The selected region is indicated by
the dashed box.

Figure~\ref{F:ccd97} shows the stars observed in the 1997 observing
season. The lines and points are as in Fig.~\ref{F:ccd96}. These stars
were primarily selected to lie outside of 3\arcmin. We observed 162
new stars based on their Washington colors and 152 yielded useful
velocities.  Of these stars, 111 (73\%) proved to be members.  Outside
3\arcmin, we measured velocities for 140 stars.  Of these, 99 (71\%)
were members. By contrast, for  stars observed in 1996 defined the same
way, 
($r>3\arcmin$ and with subsequent WIYN photometry),
only 59 of the 142 stars (42\%) proved to be members, although 
they were, admittedly, selected on the basis of an inferior 
CMD. Nonetheless, the 1997 sample extended to a larger distance from
the cluster center and would have had a lower success rate had the
candidates been selected in the CMD alone. 
Figure~\ref{F:cmd} shows the $M$ vs.\
$M-{T2}$ CMD with members
and non-members distinguished.  Most of the stars in the 1997 sample
(Fig.~\ref{F:ccd97})
have $M>15.7$ and $M\!-\!T_2<1.2$ in this CMD, while most of the
stars with $M<15.5$ are from the 1996 sample (Fig.~\ref{F:ccd96}).
A comparison between
the two samples makes it clear that member selection is easier in
the Washington color-color diagram.  

\subsection{Astrometry}

We calculated the astrometric positions of our stars using version 1.2
of the HST Guide Star Catalog (GSC) as our astrometric system
\citep[see][]{las90}.   About 9000 stars in  $45\arcmin \times
45\arcmin$ region around M92 from the `Quick~$V$'' Palomar Schmidt
survey were used as secondary astrometric standards. Plate solutions
were calculated for each of the DDO 51 WIYN frames in the mosaic; the
narrow-band filter
avoids any problems with differential refraction. The astrometric
solution was good to better than the $0\farcs 3$ required for Hydra.

\subsection{Velocity Observations} 

We obtained all observations with the same spectrograph configuration
we used for M15 \citepalias{dru98}. We used the echelle grating with
an order centered at 515 nm, in the neighborhood of the Mg~b lines.
Approximately 20~nm of the order was imaged on the 2048 pixel long CCD
for a dispersion of about 0.01 nm/pixel.  The comparison source was a
Th-Ar lamp. The Hydra multi-fiber positioner and WIYN bench
spectrograph were used.  The positions for the sky fibers were
determined from the Quick~$V$ frame.

We observed M92 over the course of four observing runs in 1996 May,
1996 June, 1997 June, and 1998 June.  Table~\ref{T:Obs} contains a log
of the observations including the number of stars observed in each
configuration. The 1996 May runs observed fewer stars than the later
runs since the proper motion list of \citet{ree92} was confined to the
inner 13.3\arcmin\ of the cluster.  The Schmidt images and the WIYN
data covered a much larger area, allowing us to observe more stars per
configuration. As the 1997 June run progressed, we did a rough
reduction of each night's data and removed stars showing grossly
different velocities from that of the cluster. This allowed us to
concentrate on the stars more likely to be members. Most of the
removed stars were in the outer part of the cluster, so, since it is
harder to place fibers in the more crowded inner region, 
the number
of stars observed per configuration went down as each run progressed.
The
proportion of members increased, however. In 1998 June we observed one
configuration in order to refine our velocities of a number of stars,
but more particularly to obtain additional spectra of a few stars with
ambiguous membership. 

In all we obtained 1204 useful spectra of 596 stars. Of these 306
turned out to be definite members as discussed below.  We include in
this total velocities for six faint stars for which we could measure
only a single velocity by combining all of their spectra. Five of
these turned out to be members. For two of these members an
additional, confirming, velocity was obtained in 1998 June.

\subsection{Velocity measurement}

We observed three stars, IV-10, III-13 and 220 repeatedly. IV-10 was
our primary standard and was stable, as was 220, but III-13 showed the
classic jitter of bright giants. We used IV-10 as our velocity
template for the velocity determination by cross-correlation.

The data were reduced using the \emph{dohydra} reduction package in
IRAF\footnote{IRAF is distributed by the National Optical Astronomy
Observatories, which are operated by the Association of Universities
for Research in Astronomy, Inc., under cooperative agreement with the
National Science Foundation.}. Each observation was accompanied by one or more 5-minute
exposures of an incandescent lamp (a ``flat'') taken with the fibers
in the same configuration as the observations.  Generally,
configurations
observed at the ends of the night had multiple flat exposures, but,
because of  the overhead involved with flats and especially with
reconfiguring the fibers, we usually only took a single flat
exposure. There did not appear to be any disadvantage to using single
flat exposures since cosmic rays were not a great problem. The program
exposures and bracketing Th-Ar lamp exposures were extracted and then
divided by the extracted lamps. In the two 1996 runs and the 1998
observation no sky subtraction was required since they were observed
in dark time. Our 1997 June run took place in brighter conditions and
some sky subtraction was required.  The sky contribution was small and
only affected the observations at the beginning of each night.

Cosmic ray removal was done through the simple expedient of using the
IRAF \emph{continuum} task to replace with the continuum fit all pixels
more than 4 standard deviations above the fit. Unlike in
\citetalias{dru98}, we did the cosmic ray removal \emph{before} the
dispersion correction and resampling. This prevented the negative spikes
we saw in the M15 data. 

As in our M15 observations, the wavelength calibration was done using
34 to 36 comparison lines.  The fifth-order dispersion solution
generally had RMS residuals of less than $10^{-4}$ nm or 0.05 km
s$^{-1}$. During dispersion correction the spectra were re-sampled
into 2048 logarithmically spaced bins covering 20.7 nm in total.

In 1996 May, 1997 June, and 1998 June we took multiple exposures in each
configuration. The resulting spectra for each star from each
configuration were summed together to produce the final spectra for
cross-correlation. The velocities were calculated using the \emph{fxcor}
task in IRAF. As in \citetalias{dru98}, we excluded all spectra with
cross-correlation peaks smaller than 0.2. These had low signal-to-noise
ratios and gave highly discordant velocities in most cases. The
template
consisted of the sum of all exposures of star IV-10 in 1996 and 1997
suitably shifted to a common heliocentric velocity. This spectrum is the
result of 28 hours of total integration time. 

Our error analysis proceeded as in \citetalias{dru98} following the
method of \citet*{pry88}.  We used the repeated observations of stars to
establish the value for the constant in the \citet{ton79} formula for
the velocity uncertainty. In this instance we obtained $C=12.8\pm 0.5$
\kms.  This is quite consistent with the value of $13.1\pm0.5~\kms$ we
obtained for M15 in \citetalias{dru98}. As the experimental setup is the
same, and M15 and M92 have similar metallicities, this is not
surprising. 

The velocity zero point was established using 18 spectra of the
twilight sky taken during various observing runs. During our 1997 June
run sky observations were taken both at evening and morning
twilight. We extracted the spectra in the usual manner and
cross-correlated them against the template spectrum. The individual
velocities were examined for fiber-to-fiber variations.  Nothing 
significant
was seen. Using the value of $C$ above, the individual
velocities have errors of 1 \kms. This is primarily an effect of the
mismatch between the low-metallicity template and the solar spectrum
of the twilight sky.  The mean velocity in each image had a standard
deviation of only 0.1 to 0.2 \kms. These means were used in estimating
the zero point.

\begin{figure}
\plotone{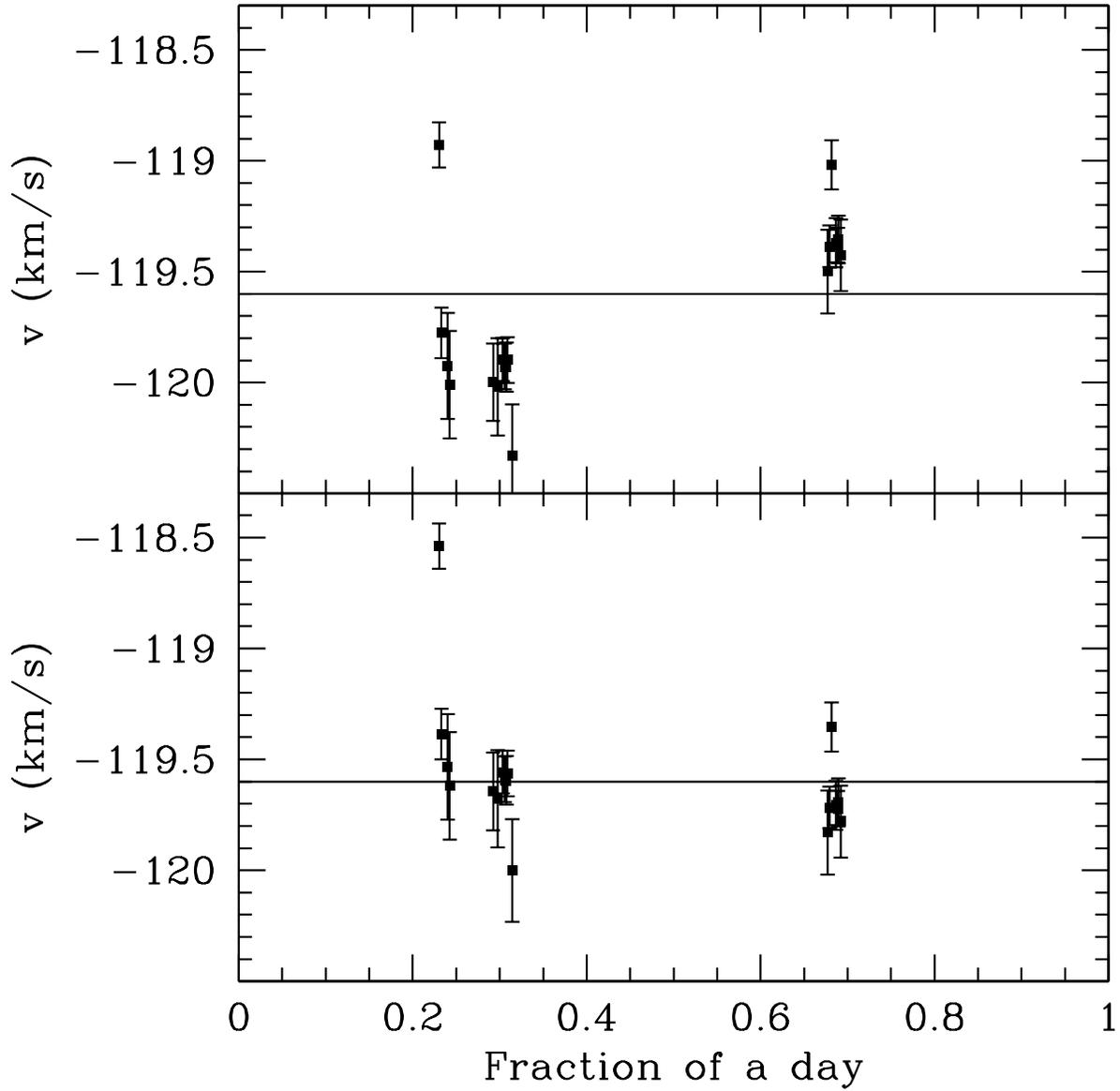} 
\figcaption[f4.eps]{(top) Relative velocity of
twilight sky exposures corrected for all motions except rotation of
the Earth plotted against fraction of a day from mean noon at Kitt
Peak. (bottom) After correction for terrestrial rotation.
\label{F:sky}}
\end{figure}

We noticed a systematic difference between the 1997 June evening and
morning exposures of about $0.54\pm 0.02$ \kms, excluding one
anomalous value. We attribute this to the rotational velocity of the Earth,
which, at the latitude of Kitt Peak and for the times of the
observations, amounts to 0.64 \kms. Subtracting this contribution
leaves a difference smaller than the errors in the means for the two
twilights. We show the results in Fig.~\ref{F:sky}. The upper panel
shows the velocities corrected for all motions except the rotation of
the Earth against the fraction of a day from mean noon at Kitt Peak.
Therefore, we corrected all the sky velocities to the heliocentric
frame. There were still some residual differences between evening and
morning observations, as seen in the lower panel of
Fig.~\ref{F:sky}. These were 0.07 \kms\ from their mean in 1997 June,
equivalent to about 0.02 of a pixel at our dispersion. The highly
consistent behavior leads us to suspect some remaining systematic
effect we are unable to identify. We are similarly unable to account
for the large deviations from the mean of a couple of
observations. Our zero-point is taken as the unweighted mean of all
the corrected sky observations and is $-119.6\pm 0.1 \kms$.

\begin{figure}
\plotone{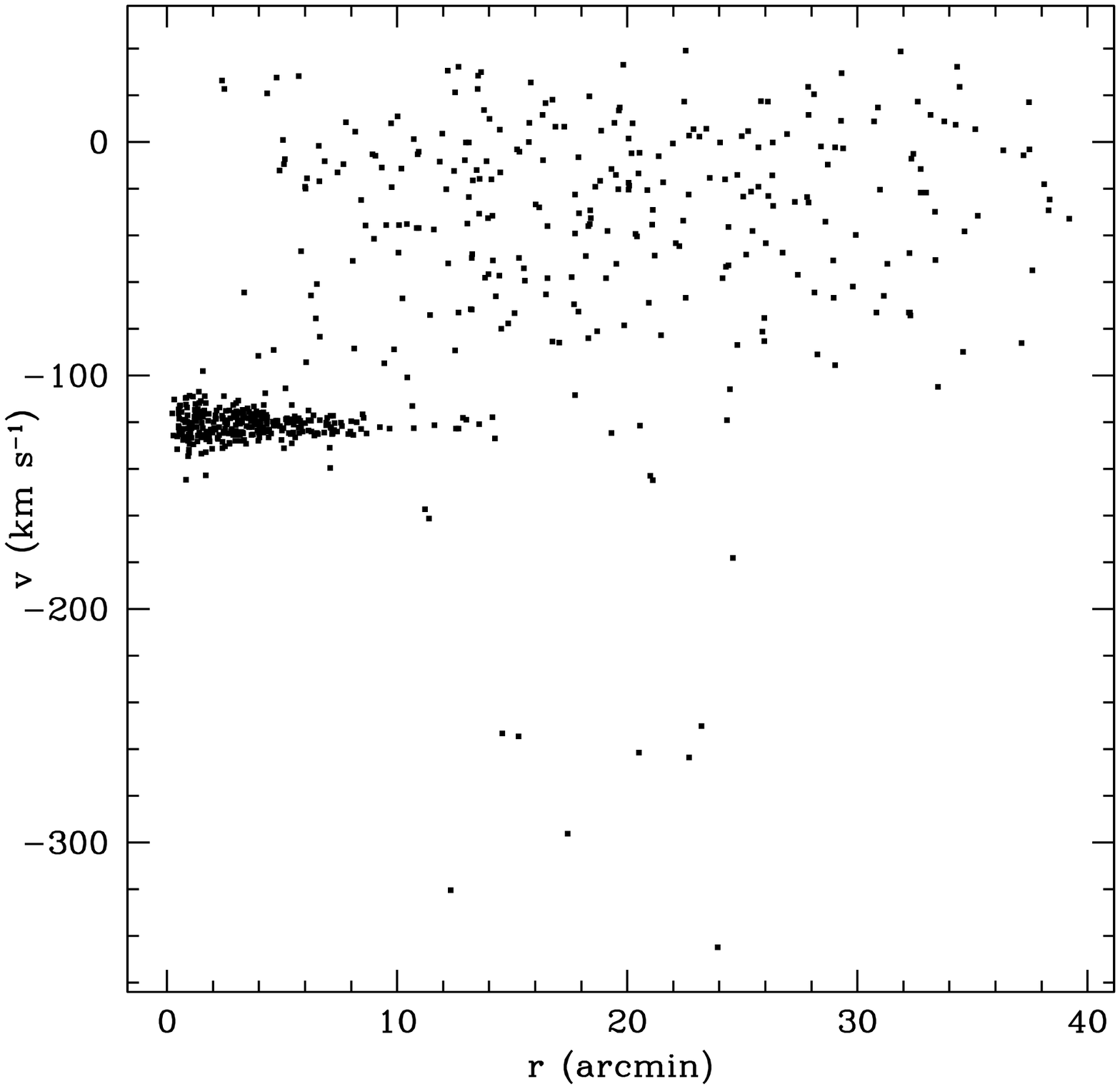} 
\figcaption[f5.eps]{Radial velocity vs. distance from
the center of the cluster for all observed stars. The cluster near
$-120$ \kms\ are mostly cluster stars.  A few presumable halo stars
are visible at more negative velocities.
\label{F:vvsr}}
\end{figure}

Figure~\ref{F:vvsr} shows the velocities for all the stars
observed. The group near $-120$ \kms\ are principly cluster members; the
Galactic disk stars cluster near zero. As in M15 we see a few stars
with
velocities significantly more negative than the cluster. These are
presumably field halo stars.

\begin{figure}
\plotone{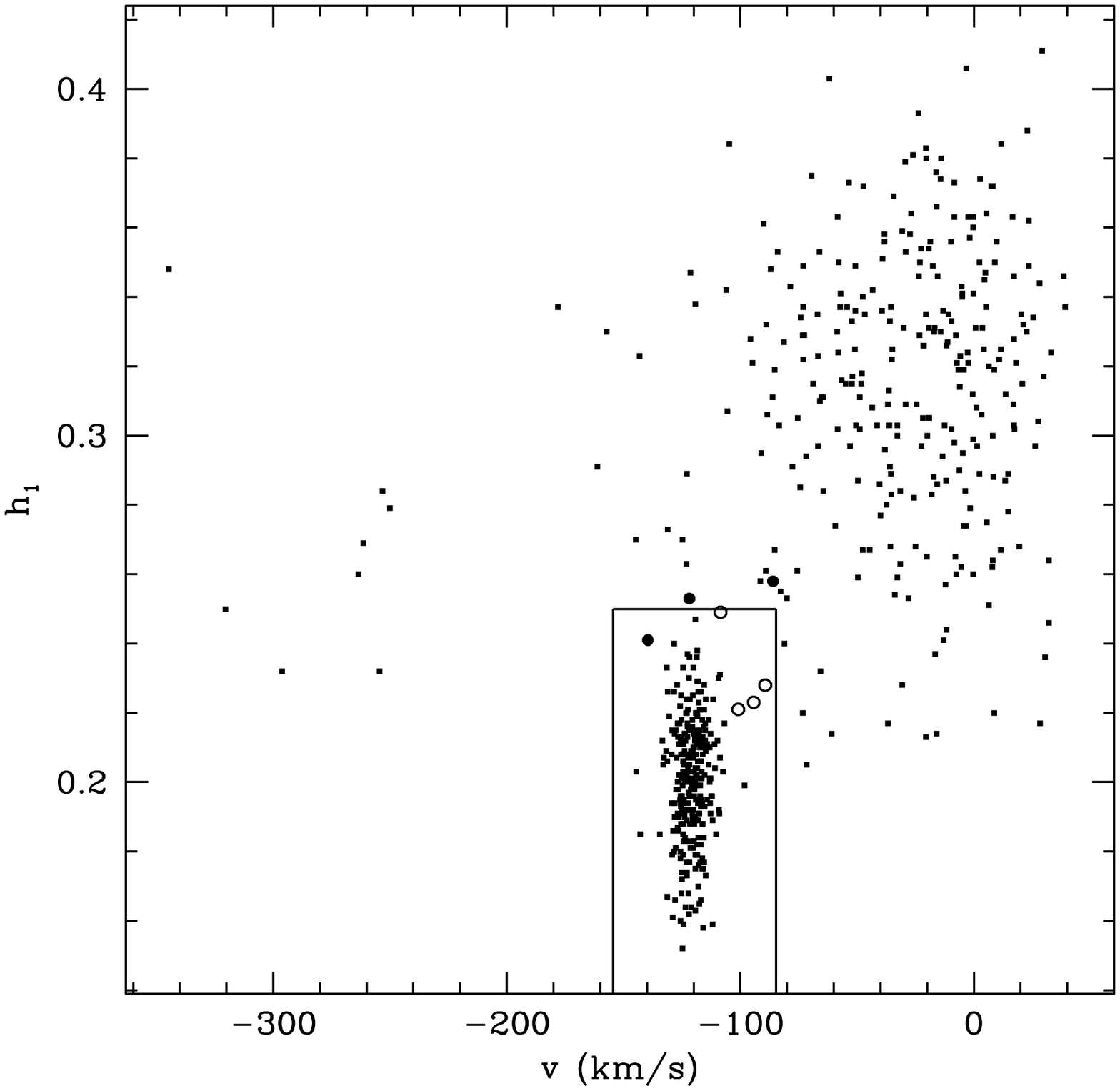} 
\figcaption[f6.eps]{Normalized autocorrelation
height, $h_1$, plotted against velocity. The box contains the locus of the members.
The open circles are the four stars rejected for the reasons
discussed in the text. The filled circles are other doubtful
members. 
\label{F:h1}}
\end{figure}

We established membership by requiring velocity coherence and line
strengths appropriate for low-metallicity giants. In \citetalias{dru98},
the metallicity criterion was simply an estimate of the equivalent width
of the Mg line at 518.3 nm. For our M92 observations, however, there
were several stars with weak spectra for which this technique gave
ambiguous results. Instead, we have used a variation of the method of
\citet{rat89}.  They showed that the height of the auto-correlation
function for a spectrum is related to the star's metallicity. Since we
simply desire an uncalibrated measure of the metallicity, we have
simplified their procedure somewhat. We have filtered the spectra in the
Fourier regime to remove the highest frequencies before
autocorrelation and have normalized the autocorrelation height by the
log of the mean number of counts in the spectrum.  We plot in
Fig.~\ref{F:h1} this $h_1$ index against the radial velocity of the
stars. 

We make the following cuts in Fig~\ref{F:h1} to establish our
members. First, we reject all stars with velocities differing from the
template by more than 35 \kms. Second, we reject all stars with
$h_1>0.25$. This leaves us with 310 stars within the box in the
figure.  Of these, four stars (open circles in Fig~\ref{F:h1}) have
been rejected from consideration for the dispersion analysis. All show
highly discrepant velocities compared with other stars at the same
distance and all but the last have zero probability of membership
based on \citet{tuc96} proper motions.  Other stars with zero
probability have been included in the sample if they otherwise satisfy
the criteria above. 

The remaining 306 stars are listed in
Table~\ref{T:members}.  Our J2000.0 coordinates are given in columns
(1) and (2), followed by a stellar identification. Where possible, we
have used identifications which are generally derived from Table 3
of \citet{tuc96}, although we have made some additions and
corrections where we have deemed them appropriate. These identifiers
primarily come from 
\citet{sw66} (hyphenated Roman numeral prefixes or `x'),
\citet{cud76} (`C'),  and \citet{bu83} (`Bu'\footnote{Stars identified
by `Bu III-nn' are those listed in Table III of that paper. All others
are from Table II.}),  with additional stars from 
 \citet{barnard} (`B'), \citet{nas38} (`N'), \citet{zinn} (`ZNG'),
\citet{saw73} (`V'), and \citet{ree92} (`R').
Where available, other names from these sources are given in
column (11).Where no previous identification exists we have give our
own number. The next column contains the distance from the cluster
center, which we take as $\alpha(J2000.0) = 17^{\rm h} 17^{\rm m}
07\fs 02$, $\delta(J2000.0) = 43\degr 8\arcmin 11\farcs 4$. The next
three columns give our Washington photometry. Column (8) gives the
number of velocities for the star. Where there was only a single
observation, the identification of the configuration in
Table~\ref{T:Obs} is given. The two stars with configuration `Z' have
single velocities based on sums of all the available spectra. Columns
(9) and (10) give the weighted mean velocity for each star and their
uncertainties. Finally, notes for the individual stars are given in
column (12). Table~\ref{T:doubt} contains the information for the
doubtful
stars and Table~\ref{T:non} (available on-line only)
contains a similar list
of non-members.

Twenty-six of the members are flagged as variables for various
reasons.  For 15, multiple radial velocity measurements show them to
be radial-velocity variables.  This is defined as having probabilities
of less than 1\% that their $\chi^2$ are consistent with no
variation. Eight stars, including two of the velocity variables, are
identified in the lists of \citet{ree92} or \citet{tuc96} with
variable stars from \citet{nas38} or \citet{saw73}.  Most of these are
RR Lyr variables.

A further group have been flagged due to the velocity ``jitter'' seen
in the brightest giants \citep{gg}. Of the 11 giants in Fig.~\ref{F:cmd}
in the
clump with $M<12.8$ and $M-{T2}>1.55$, 8 have multiple observations
and 6 of these are velocity variables. If we include an addition 0.8
\kms\ velocity uncertainty \citep{gg} to allow for this jitter, then
the velocities are consistent. We have removed these 11 stars from
consideration for our velocity dispersion analysis.

\begin{figure}
\plotone{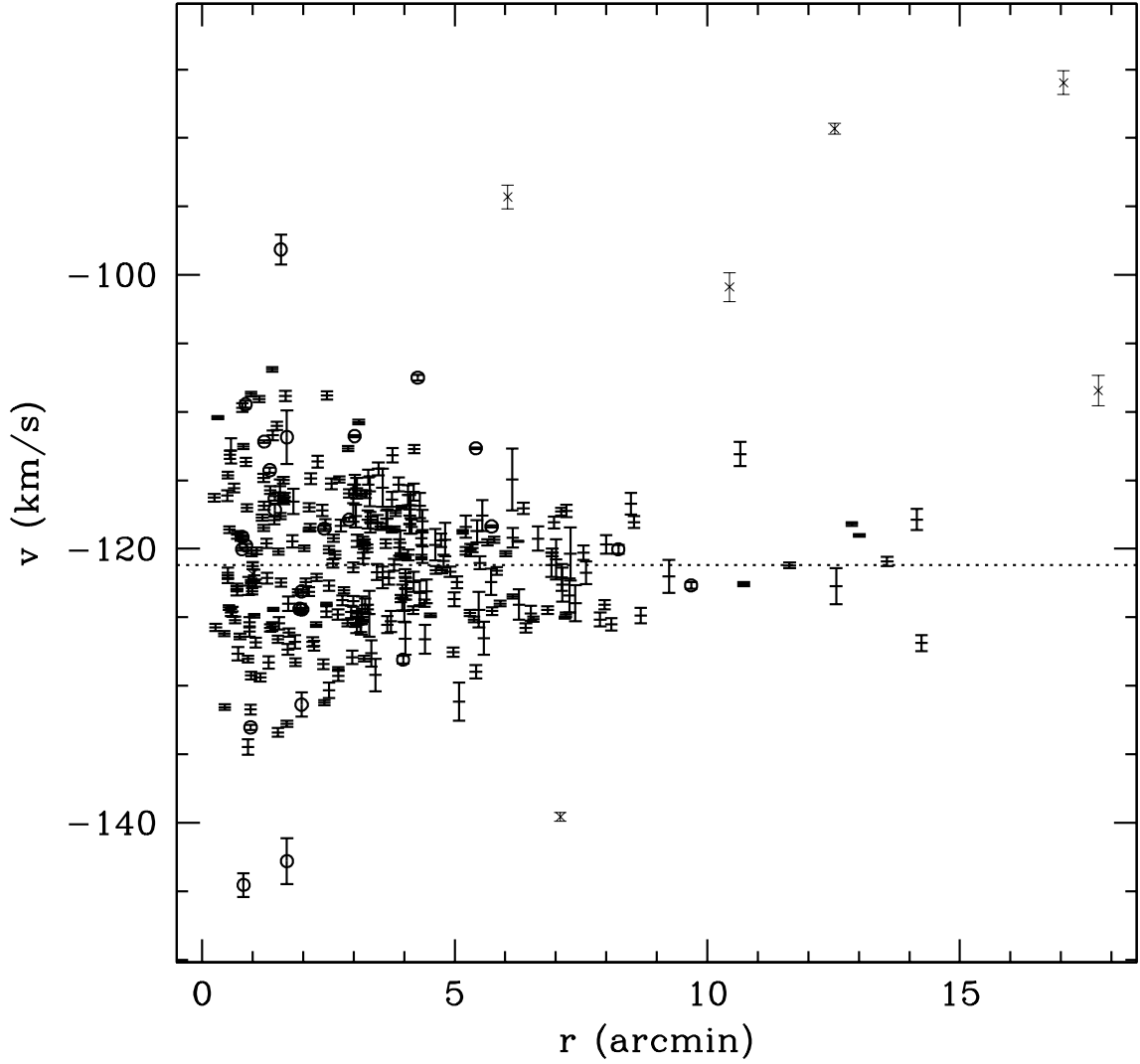} 
\figcaption[f7.eps]{Velocties vs. radial%
distance
for the members of M92. Circles represent mean velocities of the
variables. 
Note
the existence of stars with large velocities at large radii. The line
is the mean velocity. The lighter `x's are the doubtful members. (The seventh
object is in the clump near the mean velocity at 1\arcmin.) 
\label{F:members}}
\end{figure}

In addition to the four discrepant stars, we also include in
Table~\ref{T:doubt} three further stars. The stars 1016 and R644 have
accordant velocities, but slightly high $h_1$ values, and have been thus
rejected from the membership list.  On the other hand, star 1016 has
an 81\% probability of membership according to \citet{tuc96} and R644
a 99\% probability of membership according to \citet{ree92}. It is
probable that they really are members, but we have excluded them from
the analysis for consistency.  Finally, the star VI-7, which is the
star at 7\arcmin\
with velocity $-140~\kms$ has been rejected based on its peculiar DDO
colors in \citet{nor77}.

The final result is that we have 306 stars we consider to be members
and 7 stars which are doubtful members.  We show the velocities of
theses stars against their radial position in
Fig.~\ref{F:members}. The open circles indicate the 26 probable
velocity-variables mentioned above, while the fainter points are the
doubtful members. The median error for these velocities is 0.35 \kms\
with 92\% having errors less than 1 \kms.

Our sample of 306 stars includes 34 of the 35 stars for which
\citet{sod99} have reported velocities.  Comparison of our velocities
with theirs indicates a very similar mean velocity  for the
stars in their sample.  However, the individual velocity differences
from the mean show a systematic deviation between the two studies,
viz.\ their values for this difference are about 75\% of what we
find.    Their template was constructed from an average of all their bright
   giants and includes strong interstellar Na D lines as well
   (Pilachowshi, private communication). To the extent that a stellar
   spectrum is dominated by these interstellar lines, its
   cross-correlation velocity will be pulled toward the template
   velocity, i.e.\ to the cluster mean.  Thus, the measured
   cross-correlation velocity will be a weighted average of the
   cluster mean and the true velocity of the star.  Since our spectral
   region does not contain strong interstellar lines, our velocities
   are likely to be more reliable.

\section{Bayesian Analysis} 
\label{analysis}
\subsection{General Principles}
As in \citetalias{dru98}, we derive the velocity dispersion profile from our
velocities using Bayesian principles \citep{Sivia, Jaynes, MacKay}. Our
method here differs from,
and improves upon, that used in \citetalias{dru98}.  The improvements are
procedural and conceptual, as will be discussed below.

Our approach is to consider a set of specified, parametric, models for
the velocity dispersion profile. These do not exhaust the full range
of possible profiles, but do provide a plausible starting set. We then
infer the posterior probability distributions for the model parameters
for each model on the basis of our observed velocities.  We also
calculate the relative likelihoods of the various classes of models
with respect to one another. This allows us to identify the model best
supported by the data.

Generically, if the vector $\bf p$ represents the parameters for
a model or hypothesis $H$, and $D$ represents the available data, then,
given background information $I$, Bayes' Theorem states that the
posterior probability for $\bf p$ is given by
\begin{equation} p({\bf p}|D H I) = {p(D|{\bf p} H I) p({\bf p} | H I) \over p(D|H I)}. 
\end{equation}
The quantity $p(D|{\bf p} H I)= L({\bf p})$ is the likelihood
which, for the present problem, we take to be
\begin{equation} 
L({\bf p}) = (2\pi)^{-N/2} \prod_{i=1}^N k_i^{1/2} \exp\left[-{1\over 2}\sum_{i=1}^{N}{(v_i-\bar v)^2k_i}\right] 
\end{equation} 
where, if
$\epsilon_i$ is the uncertainty in velocity $v_i$ and $\sigma(r_i)$
is the dispersion appropriate to a star  at radius $r_i$, 
\begin{equation} 
k_i \equiv \left[\epsilon_i^2+\sigma(r_i)^2\right]^{-1}. 
\end{equation} 
The connection with maximum likelihood methods is discussed below. 

The components of the parameter vector are the mean velocity, $\bar v$, and the
various parameters required to define $\sigma(r)$. The mean velocity is a nuisance
parameter in terms of inferring the velocity dispersion profile, 
and can be removed by marginalizing over it. 

The prior probability for $\bf p$ is given by $p({\bf p} | HI)$.
For the most part the  parameters for the models we use are
logically independent, so $p({\bf p} | H I)$ will factor. The
denominator, $p(D|HI)= \int p(D|{\bf p} H I) p({\bf p} | H I) 
d{\bf p}$, gives the overall probability of the data given a
hypothesis which is variously termed the `evidence' or the `model
likelihood'.
 For a particular model this is just a normalization
factor, but application of Bayes' Theorem once again gives $p(H|DI)
\propto p(D|HI) p(H|I)$ for the posterior probability of model H.
When we wish to compare two models, $H_1$ and $H_2$ we can do this
by taking the ratio of their posterior probabilities:
\begin{equation}
{p(H_1|DI)\over p(H_2|DI)} = {p(D|H_1I)\over p(D|H_2I)}{ p(H_1|I)\over p(H_2|I)}.
\end{equation} 
The first factor on the right hand side is the likelihood ratio using the values of
$p(D|HI)$ we just computed. The second factor is the ratio of the
prior probabilities for the two models. Our procedure in \citetalias{dru98}
had the shortcoming in that instead of calculating $p(D|HI)$,
we compared the maxima in $p({\bf p}| D H I)$. This compares only
the best set of parameters for each model. The proper thing to
consider is the relative merits of the models for all  combinations
of parameters. Comparing the probabilities of the best sets of
parameters favors a model having a narrow range of parameters that
fit the data particularly well over one that has a broader range of
parameters that fit the data somewhat less well, whereas the
opposite ought to be the case. The issue is the choice of the best
model class,
not simply the best set of parameters. 

In this we differ from maximum likelihood methods. The latter can be
derived from the Bayesian method in the limit where the prior
probabilities are constant. The objective is then to find the maximum,
i.e. the mode, of the likelihood function and to report the function
based on 
this set of parameters as ``the velocity dispersion profile'', in much
the same way as we did in \citetalias{dru98}. The width of the
 likelihood function can be used to give uncertainties for the
parameters, but this method has no way of assessing the relative
merits of 
various possible descriptions of the data. 

At the same time, it is often difficult to  illustrate fully the
complexity of the posterior probability function. 
For situations with a large number of parameters, Markov-chain Monte-Carlo
methods are practical, but can take a long time to
generate sufficient samples to get ensure that the posterior is
being properly sampled. For problems with
relatively few parameters, direct calculation
of the probability is a practical proposition. Consequently, in this
paper, we compute the joint probabilities $p({\bf p}| DHI)$ directly
on a multi-dimensional grid. The highest dimensional grid we used here
has five dimensions, but the number depends on the model as discussed
in the next section. For models with radial binning, the parameter
space naturally factors into smaller, tractable, sub-spaces. In each
dimension we used a grid of 50 points concentrated on the region of
significant probability and is well resolved. 
We then extract the information in this function in several ways to
highlight not only its maximum, but also certain aspects of its shape.

To this end we calculate the marginal probabilities, $p(p_i|DHI)$, of
the each of
the parameters, $p_{i}$ by integrating $p({\bf p}| DHI)$ over all the
other parameters in the problem.
Numerically, we present our results in terms of the moments and mode
of these marginal distributions. For each parameter we give the mean
defined as the expectation value
\begin{equation}
\langle p_{i}\rangle \equiv \int p_{i}p(p_{i}|DHI)dp_{i},
\label{E:mean}
\end{equation}
and the standard deviation
\begin{equation}
\epsilon_{p_{i}} = \sqrt{\langle p_{i}^{2}\rangle -
\langle p_{i}\rangle ^{2}}.
\label{E:sigma}
\end{equation}
We also estimate the mode of the smooth marginal
distributions by constructing a parabolic fit to the mode  and
the two bracketing points on the
function grid. Deviations between the mode and the mean
highlight asymmetries in $p(p_{i}|DHI)$. In such cases 
we will use the following notation to indicate both these values:
\moments{x}{\langle x \rangle}{x}{\epsilon_{x}},
where the value in parentheses is the mode. Otherwise, we will just
give the mean and $1\sigma$ standard deviation as usual, although
this should not be taken to necessarily indicate that the marginal
distribution is Gaussian.

It should be noted that the modes of the marginal distributions
correspond to the overall mode of $p({\bf p}| DHI)$ only if the  joint
probability is symmetric. In practice, there are often significant
correlations between parameters as highlighted by covariances
calculated from joint probabilities of pairs of parameters. In order
to highlight the range of parameters permitted by $p({\bf p}| DHI)$ we
give two graphical representations of our results in addition to
the values mentioned in the previous paragraph. First, we will show
the marginal probability distributions for the best-fitting models. We
also calculate $p(\sigma|rDHI)$ by integrating
$p({\bf p}| DHI)$ over regions of constant $\sigma(r)$. From these we
will show the
mean value and uncertainty as defined by equations (\ref{E:mean}) and
(\ref{E:sigma}), as well as the mode. Again, the difference between
the mode and mean highlight asymmetries in  $p(\sigma|rDHI)$. In
practice, these curves will also deviate from those based on the mean
or modal parameters for the model, again because of asymmetries in the
overall probability distribution. For ease of graphical comparison
with later models, we shall tabulate some of the points on these
curves. We wish to emphasize that the errors on these points are by no
means independent.  Comparisons between future models and these
data are best undertaken with the original velocities, not with
these curves.

In light of these new, improved, procedures, we will also reanalyze the
M15 data from \citetalias{dru98}. 

\subsection{The Models}
\label{models}
In accord with the discussion in \cite{dru03} highlighting the
underlying
difficulties in inferring global properties from a set of individual
velocities, the velocity-dispersion profiles to be derived below
depend on the following assumptions. We assume that the velocities can
be perfectly described by Gaussian distributions with a common mean
and a dispersion that can vary with radius. Any higher-order moments
will be ignored in our analysis. We assume that the uncertainties
derived above for our radial velocities are accurate and precise and
that the noise in the measured velocities are distributed as Gaussians
with zero mean and standard deviation equal to these uncertainties.
We assume that there are no non-members present in our sample. We do
not exclude the possibility of there being unbound stars. Indeed, the
use of a Gaussian implicitly assumes that the extreme tail is
populated by unbound stars. We also assume that all the stars in our
final sample are single stars with the same mass. It is possible that
there are unidentified
velocity variables, including binaries, in our sample. These would serve to 
inflate the dispersion. These provisos are included in the
background information $I$.  One model which could be considered is
one where the distribution of the velocities is consistent with the
observed errors. Such a model, however, is strongly rejected with
respect to all others, indicating that we detect the intrinsic variance
in the cluster velocities. We now list the models we will consider in
this paper. This list is not exclusive, nor exhaustive, but does cover
likely variations.\footnote{Which is not to preclude other,
potentially interesting models. The \citet{k66} model, has often been
used to describe globular clusters. With its energy cut off,
however, it explicitly cannot account for escaping stars, which
would then have to be accounted for separately. Such two-component
models, as discussed at the end of this section, are beyond the
scope of this paper. Similarly, we do not investigate the
models including non-Newtonian effects
 \citep{scar03,scar06}.}
 
Model $B$: The stars are sorted by distance from the cluster center
and divided into bins. The velocity dispersion is taken to be
constant in each bin. This is the traditional approach,  which we
augment with our Bayesian estimation of the velocity dispersion.
The parameters are the mean velocity of the
sample, $\bar v$, the number of bins $M$, and the set of dispersions,
$\{\sigma_j\}$, $j=1,\ldots, M$. The choice of binning leads
to various sub-types of this model class.  Bins can be arbitrary, or
divided
roughly equally in terms of number of stars per bin or in terms of the
radial range covered. An alternative is to let the numbers of stars
per bin, or equivalently the radial ranges, be an additional set of
$M-1$ parameters, ${r_{i}}$, $i=1,\ldots, M-1$.

Model $P$: The velocity-dispersion profile has the form of a power-law
$\sigma(r)=\sigma_1 r^\alpha$. The parameters of this model are:
$\bar v$; $\sigma_1$, the velocity dispersion at $r=1\arcmin$; and $\alpha$.

Model $C$: The velocity-dispersion profile has the form of a power-law
with a core:
\begin{equation}
\sigma(r) = \sigma_1 \left\{{1+\left({r/ r_0}\right)^2\over 1+r_0^{-2}}\right\}^{\alpha/2}.
\end{equation}
The parameters are as in model $P$ with the addition of the scale radius
$r_0$ of the core in arc minutes. Note that, as in model $P$,
$\sigma_1$ is normalized to
be the value of the dispersion at $r=1\arcmin$. This removes the correlation
between $r_0$ and $\sigma_1$ that otherwise would be present. Model $C$
is identical with model $P$ in the limit $r_0\rightarrow 0$, and so is a
more general case. For large $r$, $\sigma \propto r^\alpha$.

Model $PB$: This model is a hybrid with a power-law inner section and an
outer section where the velocity dispersion in constant. This model is
motivated by our discovery in \citetalias{dru98} that the velocity dispersion
appears to increase in the outer part of M15. It seems sensible then
to consider that these stars may not share the velocity dispersion
properties of the rest of the stars in the cluster.  The parameters
here are the same as in model $P$, with the addition of the radius $r_1$
dividing the two populations and the dispersion, $\sigma_{\rm out}$,
for these outer stars.

Model $CB$: This model is as model $PB$, but for the cored power-law. It
has the parameters of model $C$, plus $r_1$ and $\sigma_{\rm out}$.

In both models $PB$ and $CB$ the assumption is that there are distinct
radial regions containing distinct populations of stars. The inner
population are the stars bound to the cluster, while the outer
population consists of those stars which deviate from the cluster
distribution because of tidal perturbations, ejection from the
cluster core, or other, unknown reasons. A more realistic model would
be one where the two populations are intermingled with additional
parameters, including the membership probabilities for the two
populations, describing the composite distribution. Similarly, the
possibility of undiscovered binaries could be included.  These are a
more complicated propositions to implement than the models presented
here, but may be required to fully understand the data. A description
of a similar approach can be found in \S 21.5 of \citet{Jaynes}.

\subsection{Priors}
\label{S:Priors}
We now consider the prior probabilities for the
various parameters.
The mean velocity is common to all the models and we take $p(\bar v|HI)$
to be uniform.  Its normalization will drop out of all model
comparisons, so we can safely use a flat, improper prior.
 For the various dispersions, the correct prior is the
Jeffreys prior \citep{Sivia,Jaynes} $p(\sigma|HI) \propto \sigma^{-1}$. This is equivalent
to a flat prior in $\log(\sigma)$. We will need to set
sensible bounds, $\sigma_- < \sigma < \sigma_+$, on this in order to
make it normalizable. We have used $\sigma_- = 0.3\ \kms$, since this is
the median error on our velocities, and smaller dispersions are likely to
be meaningless. For the upper limit we use  values appropriate to
each
cluster. 
For the models using the power-law index $\alpha$, we have taken
$p(\alpha|HI)$ to be uniform over $-2.5<\alpha<0$ since the velocity
dispersion should be a decreasing function of radius, but not too strongly. The lower
limit is set by  consideration of the range of physically sensible values for $\alpha$.

For models $C$ and $CB$, since we interpret $r_0$ as a scale
parameter, the correct prior is the Jeffreys prior. It is meaningless
to speak of a core radius outside the radial range of the data, so the
limits on $r_0$ are taken with this in mind.

Finally, for models $PB$ and $CB$, as well as variants of model $B$
with varying bin sizes, we take $p(r_i|HI)$ to be uniform over the
appropriate radial range. In this study we will consider two different
sorts of binnings. The first is a standard arrangement, with the stars
divided into bins of equal number. The second is to set a minimum
number of stars per bin
and then scan through all possible combinations of break radii in
order to find the optimal bin size. We will denote models with these
two binnings as $B_{N=M}$ and $B_{S=M}$ respectively, where $M$ is the
number of bins.  A scanning approach similar to the $B_S$ models is
taken to the break radius $r_1$ in models $PB$ and $CB$, and, again,
the prior is uniform over a specified radial range.
 
Note that for models in class $B$ with $M>2$, the prior probabilities
on the positions of the bin boundaries are not independent, since they
are sequential and we require each bin to contain a minimum number of
stars. For $M=3$, for example, $p(r_1 r_2 |B_{S=3}I) =
p(r_1|B_{S=3}I)p(r_2|r_1 B_{S=3}I)$ follows from the product rule of
probabilities, with similar expressions for additional bin boundaries.

The limits on the various parameters for the M92
and M15 data sets are given in Table~\ref{T:range}.

The selection of priors for the models themselves is a somewhat more
difficult proposition. In some cases, we may indeed have pertinent
prior information.  For M15, for example, the surface-density and
surface-brightness profiles are power-law in form to within 2\arcsec\
of the center of the cluster \citep{sos97}.  Inspection of the
velocity-dispersion profile in Fig. 9 of \citet{ger02}, suggests that
the break in the slope of the velocity dispersion occurs somewhat
further out, at  around 1\arcmin\ to within  a factor of 2. 
This difference is to be expected. If, for example, we consider the
single-mass $W_{0}=9$ King (1966) model shown in Fig. 4-11 of
\citet{BT}, we can see that the line-of-sight velocity dispersion
drops to half its central value at about $30 r_{c}$. We could consider
this an effective core radius for the velocity dispersion profile,
$r_{c,v}$. Thus, it would not be unreasonable to expect to see some
evidence for flattening within our data sample. If the core radius in
the velocity dispersion is unresolved, it may be difficult to
differentiate between a true core, and the situation we will see in
Example {\bf PL+F} of \S\ref{S:Examples} below. Taking this into
account,
$p(P|I_{\rm{M15}})/p(C|I_{\rm{M15}})\sim 5$ would not be unreasonable
for  the two options of a power-law (model $P$) or a power-law with
core (model $C$). This expresses our relative ambivalence with respect
to the two models, but a desire for some stronger evidence of
flattening before we are prepared to believe it. 
Like any prior, it is a statement of how strongly
we hold any given proposition. 

For M92, our preference is for model  $C$ over model $P$, since the
cluster shows a clear core radius of 14\arcsec\ in the density data
\citep{trager}. We would 
then expect $r_{c,v}\sim 7\arcmin$. In terms of the parameterization
of model $C$, $r_{0}=r_{c,v}/\sqrt{4^{-1/\alpha}-1}$. For
$\alpha<-0.25$, $r_{0}>1\arcmin$, and is well within our data sample.
Thus we expect to see clear evidence for a core in the velocity data. 
In principle, this argument can lead to a stronger joint prior on
$r_{0}$ and $\alpha$ than the relatively uninformative priors proposed
above, but we shall refrain from doing so here. Instead, we take
$p(C|I_{\rm{M92}})/p(P|I_{\rm{M92}})=100$, expressing our strong prior
expectation of model $C$.

In the absence of suitable prior information, the same considerations
which lead to the numerical values above suggest we employ the
``device of imaginary results'' of \citet{good}. Before doing the
analysis of the results, we ask ourselves how much better would the
likelihood of one model have to be over another for us to consider the
two models equally probable. The inverse of this ratio is then the
desired prior. If, for example, the possible models for a set of
globular cluster data are either one with or one without a central
black hole of unknown mass, and, being skeptical of the proposition,
we would need the likelihood in support of a black hole to be 50 times
that without, then the proper prior ratio is $p(B|DI)/p(\bar
B|DI)=1/50$ where $B$ is the proposition that the cluster contains a
black hole and $\bar B$ its negation.

The case of binned models is interesting in this regard. No one
actually thinks of a globular cluster as being stratified in the way
implied by binning. The main reason to use such a model is to look for
features in the profile not described by the simpler parametric models
such as $P$, $C$, etc. Increasing the number of bins allows for finer
detail in the resulting profile, but also decreases the precision of
the measurement of the dispersions. The Bayesian formalism naturally
takes this effect into account. A model with more bins is penalized by
the additional priors attached to additional parameters. If,
nonetheless, the total probability is maximized for a particular set
of bins, this is a way to optimize the choice. Such optimization is,
of course, subject to the prior on the number of bins, but we shall
take that to be uniform up to some maximum number of bins. Our
objective with this class of models is to look for the most probable
binning in case there is some obvious feature not captured by the
other models under consideration. Nonetheless, given the lack of
physical motivation for a binned model, we set $p(B|I)$
to be 0.02 times the prior probability of the preferred
model
of $P$ or $C$ as appropriate for M92 and M15. This probability is
intended to apply to the entire collection of binned models
reflecting
the flexibility the observer has in choosing the binning. 
In the subset of models with variable-width bins we require a
minimum of 20 stars per bin for $M=2$, $3$ or $4$. (The calculation
becomes very long for higher values of $M$ in this case.)
Applying the
same limit to bins with an equal number of stars per bin, permits all
models with $M\le 14$ for the M92 data and $M\le 10$ for the M15
data set. Dividing $p(B|I)$ equally between
the 17 cases for M92 and 13 cases for M15 gives  relative prior
probabilities of $1.2\times 10^{-3}$ and $1.5\times 10^{-3}$ per
binned model for the two clusters. 

Finally, we consider the odds to be even on whether the outer zones in
models $PB$ and $CB$ exist or not. 

At this point, we invite the reader to decide  what prior
ratios
are appropriate for the various models we describe in
\S\ref{models} in the cases of M92 and M15. Below, we will present the
results in such a way that the posterior odds can be recalculated
for other choices of prior ratios.

\subsection{Implementation Details}

The posterior probability distribution for each model was calculated on
an $N$-dimensional grid, where $N$ is the number of parameters in
the model.
The marginal probability for each
parameter was calculated from the values on this grid by numerical
integration over all the other parameters. Along the way, the
two-dimensional marginalizations were examined and used to
calculate the covariances between pairs of parameters. To increase
the efficiency of the calculations, we first conducted a pilot
study using a coarse grid covering the full ranges of the priors.
This was used to locate the region of non-trivial probability in each
parameter, which was then covered by a finer grid in our final
calculation. The  marginal probability distributions were saved. We
also calculated the first and second moments of these to give the
expectation value and variance of the distributions. In cases where
the posterior probability distribution is close to a Gaussian,
these are close to the mean and variance of that Gaussian. 

In the cases where one or more break radii are themselves
parameters, the above calculation was done for each set of break
radii and the total probability is naturally the marginal
probability of that set of radii. The posterior probability
distributions of the break radii are highly non-Gaussian and rather
than specifying the rather unenlightening moments, we use the mode
as the typical value.  The other parameters were marginalized over
the break radii.

\subsection{Examples}
\label{S:Examples}

We now proceed to use some mock data sets to demonstrate the use of
our methods. In all cases, the radial distributions and velocity
uncertainties are those of our M92 data.  For each mock data sets we
used our Bayesian procedures to estimate the parameters for all the
models discussed. For the binned models we used equal number bins
running from 1 to 20 bins and scanned bins with 2 to 6 bins, i.e. with
1 to 5 breakpoints, and a minimum bin size of 40 stars. We present our
results in Table~\ref{T:mock}. The log of the likelihood ratio with
respect to the best model is listed for each of the mock data sets.
For each of the binning schemes, the total probability presented is
the sum of the total probabilities for all binnings. We also give the
results for the best value of $M$ in each binning scheme.  In all
cases, the mean velocity of the mock data sets was consistent with the
input value. Although we have dropped the units for simplicity, they
should be clear from the context. We
discuss each mock data set in turn.

{\bf PL}: A power-law with $\sigma_1 = 6.5$ and $\alpha = -1$. The
best model is $P$ with \momentsa{\sigma_{1}}{6.0}{5.9}{0.4}\
and  \momentsa{\alpha}{-1.05}{-1.05}{0.06}. Model $C$ is 50 times
worse, and, since the best core radius is at the lower limit of the
prior, this supports the primacy of model $P$ in this case. The only
comparable model is $PB$, where the $\sigma_1$ and $\alpha$
probability distributions are nearly identical. In the outer region,
$p(\sigma_{\rm out}|DI)$ is strongly peaked at the lower limit of the
allowed range. In other words, the results for model $PB$ are
effectively identical with those of model $P$, but the model is
penalized by its additional parameters. The steep gradient makes the
constant dispersion within each of the 
bins in 
model $B$ a very poor representation of the data. 

{\bf CPL}: A cored power-law with with $\sigma_1 = 6.5$ and $\alpha =
-1$ and 
$r_0=1$. The best model is $C$ with
\momentsa{\sigma_{1}}{6.2}{6.1}{0.5},
\momentsa{\alpha}{-1.0}{-0.9}{0.1}, and \momentsa{r_{0}}{1.1}{1.0}{0.4}.
The next best model is $CB$, but the situation here is the same as in
the previous case. The parameters in the inner region are the same and
the fit is penalized by the existence of the additional parameters.

\begin{figure}
\plotone{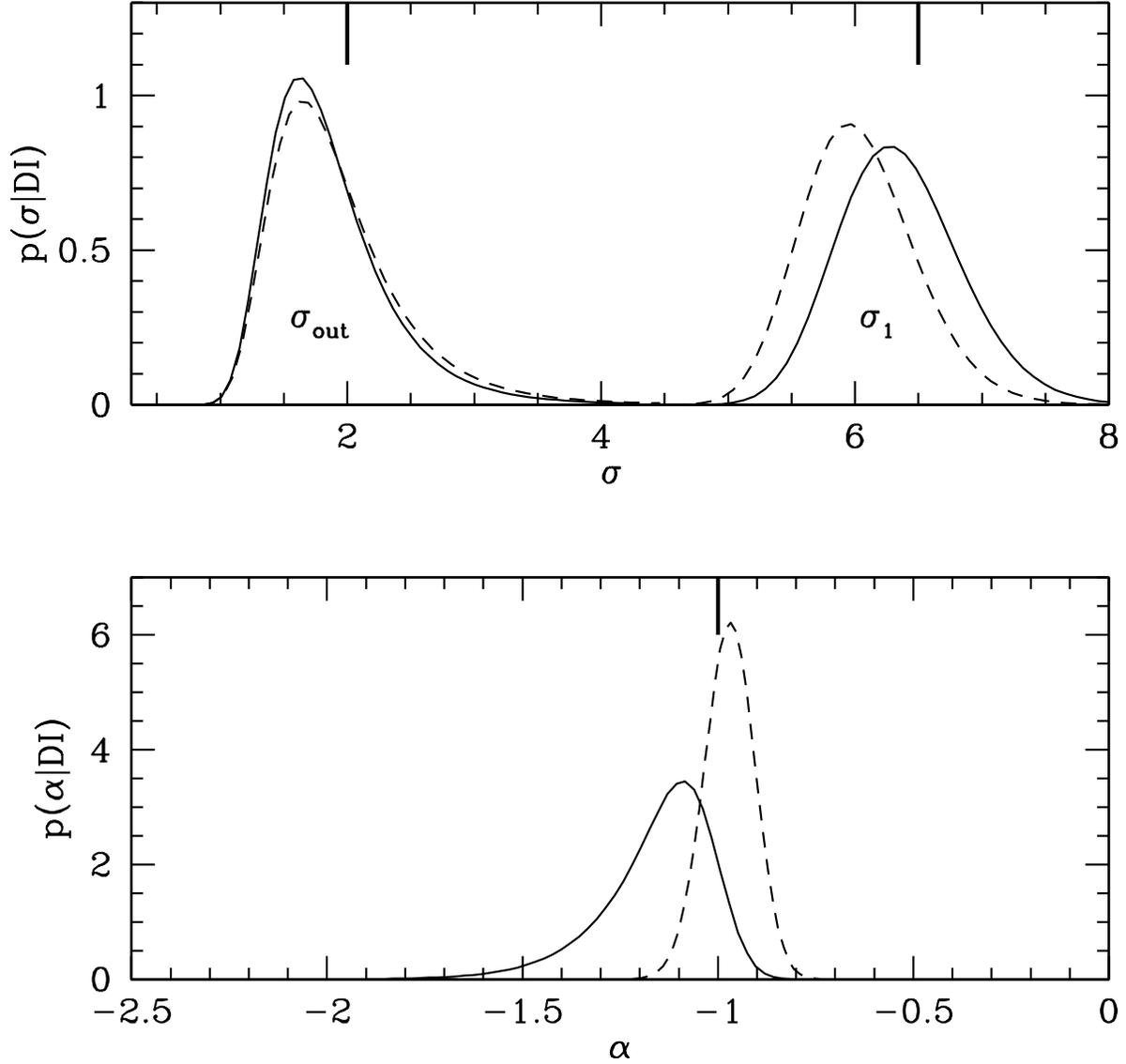}
\figcaption[f8.eps]{Probability density distributions for the
velocity dispersions (top)
and power-law index (bottom) for the mock data set {\bf PL+F}. Solid
lines: model $CB$. Dashed lines: model $PB$. The thick vertical
lines are at the  input values of the three parameters. 
\label{F:fakeF}}
\end{figure}

\begin{figure}
\plotone{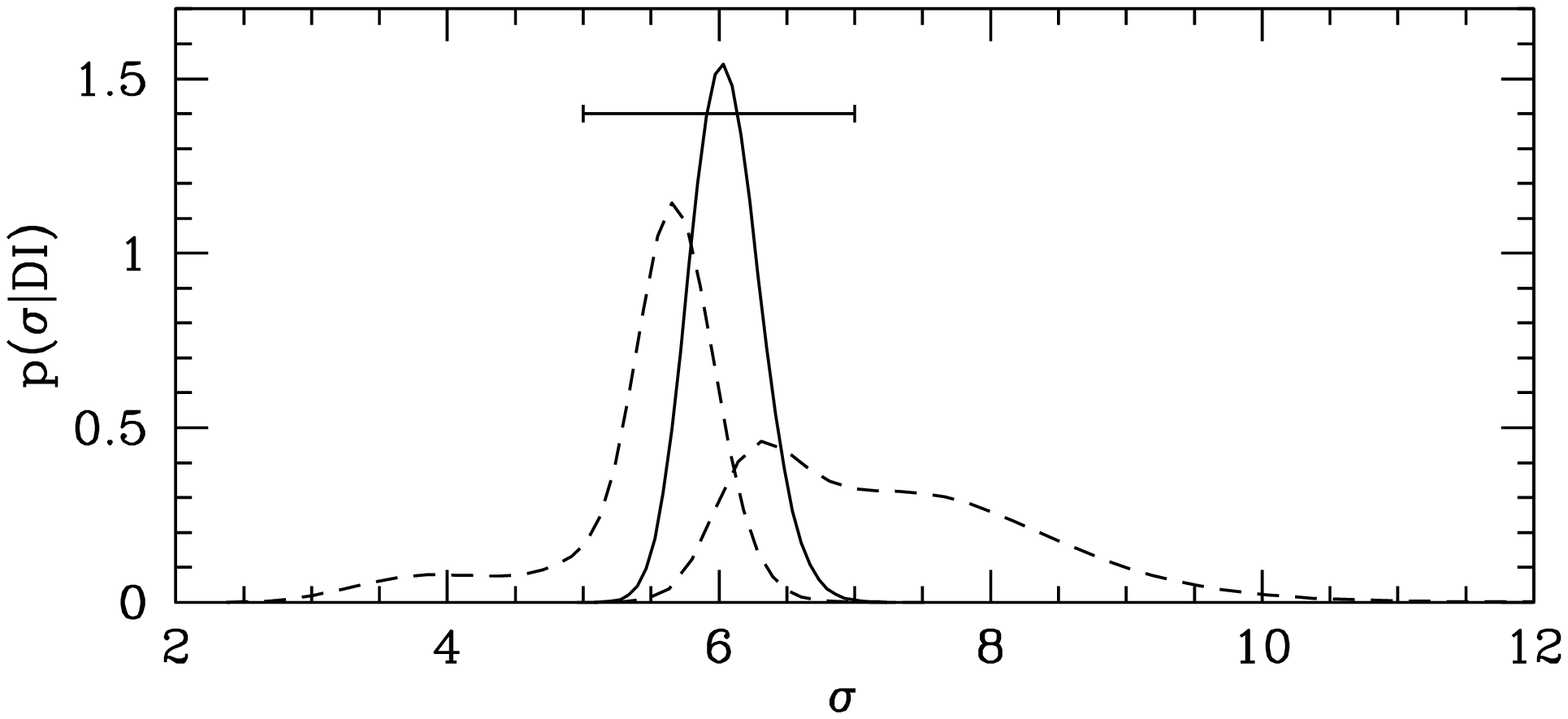} 
\figcaption[f9.eps]{Probability density
distributions for the velocity dispersion for mock data
set {\bf S}. Solid line: Single dispersion model. Dashed lines: Two-dispersion
model. The line at top represents the range of input dispersions for
this data set. 
\label{F:fakeZ}}
\end{figure}

{\bf PL+F}: A power-law with $\sigma_1 = 6.5$ and $\alpha = -1$
inside a radius of 8, but
beyond this the
dispersion is flat at 2. The most probable model is $CB$, with $PB$
running a very close
second.  The core radius probability distribution is strongly peaked
at the lower limit of the permissible range (0.3), giving only an upper
limit. This indicates some flattening of the dispersion profile
towards small radii, but the core is not resolved. This is easily
understood in terms of the spatial distribution of the stars. In the
innermost sampled region, only a small fraction of the stars have
measured velocities.  The probability of any of these stars having a
high
velocity is small, so the velocity dispersion appears to increase less
quickly at the inner edge of the data. Thus we can see hints of a core
radius, even when the model underlying the mock data contains no such
flattening.  The estimated values of $\sigma_1$ and $\alpha$ are
similar in the two models, but are less well defined in the more
general model $CB$ as seen in Figure~\ref{F:fakeF}. As is also shown,
the two models do agree on the dispersion in the outer region. The
asymmetry in the probability density distribution for $\sigma_{\rm
  out}$ is typical of dispersions measured from a small number of
stars. For model CB
\momentsa{\sigma_{1}}{6.4}{6.3}{0.5},
\moments{\alpha}{-1.2}{-1.1}{0.1},
\moments{r_{0}}{0.6}{0.3}{0.3}, and \moments{\sigma_{\rm
out}}{1.9}{1.6}{0.5}, while for model $PB$
\momentsa{\sigma_{1}}{6.0}{6.0}{0.4},
\momentsa{\alpha}{-0.97}{-0.97}{0.06},
 and \moments{\sigma_{\rm out}}{1.9}{1.7}{0.5}. There is a minor
disagreement as to the location of the boundary between the two
regions depending on which side to place the three points lying
between 8\farcm 3 and 10\arcmin.

{\bf D}: A flat distribution with $\sigma=6.5$ except between
$1\farcm
5$ and $3\farcm 5$, where the dispersion is double. This is admittedly
unrealistic, but is designed to show the flexibility of the procedure.
The best model is a binned one with two adjustable boundaries giving
three bins. The modal boundaries lie at $1\farcm 53$ and $3\farcm 56$
 and the velocity dispersions in the three bins are
\momentsa{\sigma_{1}}{7.2}{7.1}{0.7},
\momentsa{\sigma_{2}}{13.2}{13.0}{1.0}, and
\momentsa{\sigma_{3}}{7.4}{7.3}{0.5}, with the bins containing 65,
91, and 118 stars in order of radius. The next best model is one with
three adjustable bin boundaries, but the final result is degenerate in
the sense that the outer two boundaries have the same mode. This model
is 7 times less probable than the best model. The other models are at
best $10^4$ times less probable, and must be rejected for any sensible
prior. 

{\bf S}: The velocity dispersion is given by $\sigma(r)=5.+2.\sin(r\pi/14)$. 
This mock data set is designed to show what happens when none
of the models is suitable to the data. In this
case, the most probable model is simply a flat distribution with
\momentsa{\sigma_{1}}{6.1}{6.0}{0.3}. Nearly as likely is a
two-zone model with a modal bin boundary at 6\arcmin. In this model
\momentsa{\sigma_{1}}{5.6}{5.6}{0.4} and
\moments{\sigma_{2}}{7.3}{6.5}{0.9}.  As shown in Figure~\ref{F:fakeZ} the
probability distributions for this latter model are quite
non-Gaussian, with the probability distributions for both dispersions
showing hints of bimodality. Results like these for
real data would indicate a great cause for concern. 

We now turn to our velocities in M92 and M15 and analyze them in the
same way.

\section{Kinematic Results}
\label{kinematics}
\subsection{M92}
We give the results of our Bayesian analysis of the new M92 data
in Table~\ref{T:M92}. The likelihood of our five model classes relative
to model $CB$ is given in the second column of the table. With the
priors in the third column---based on the discussion in
\S\ref{S:Priors}---the posterior probability of the model classes is
given in the fourth column. Model $CB$ is the most likely, but model
$C$ is only a factor of 2 less probable, so we shall consider them
together. Models $P$ and $PB$ are an order of magnitude less likely
than mode $CB$, confirming the existence of a core, and with our prior
information, they are up to $10^{3}$ times less probable.  

Of the $B$-class models, the model with two equal sized bins is
significantly more likely than any of the others. It is interesting to
note that the model with two adjustable bins comes up with much the
same result, shuffling eight stars from the inner to the outer bin, at
the cost of a lower likelihood due to the prior on the position of the
bin boundary. The $B_{N=2}$ model has dispersions of
\momentsa{\sigma_{1}}{5.8}{5.8}{0.4} \kms\ and
\momentsa{\sigma_{2}}{3.5}{3.5}{0.2} \kms. At the bottom of
Table~\ref{T:M92} we give the likelihoods relative to model $CB$ for
the most likely binned models. Increasing the number of bins does not
improve the match to the data. Given our prior, we reject the binned
models, individually and as a class.

\begin{figure}
\plotone{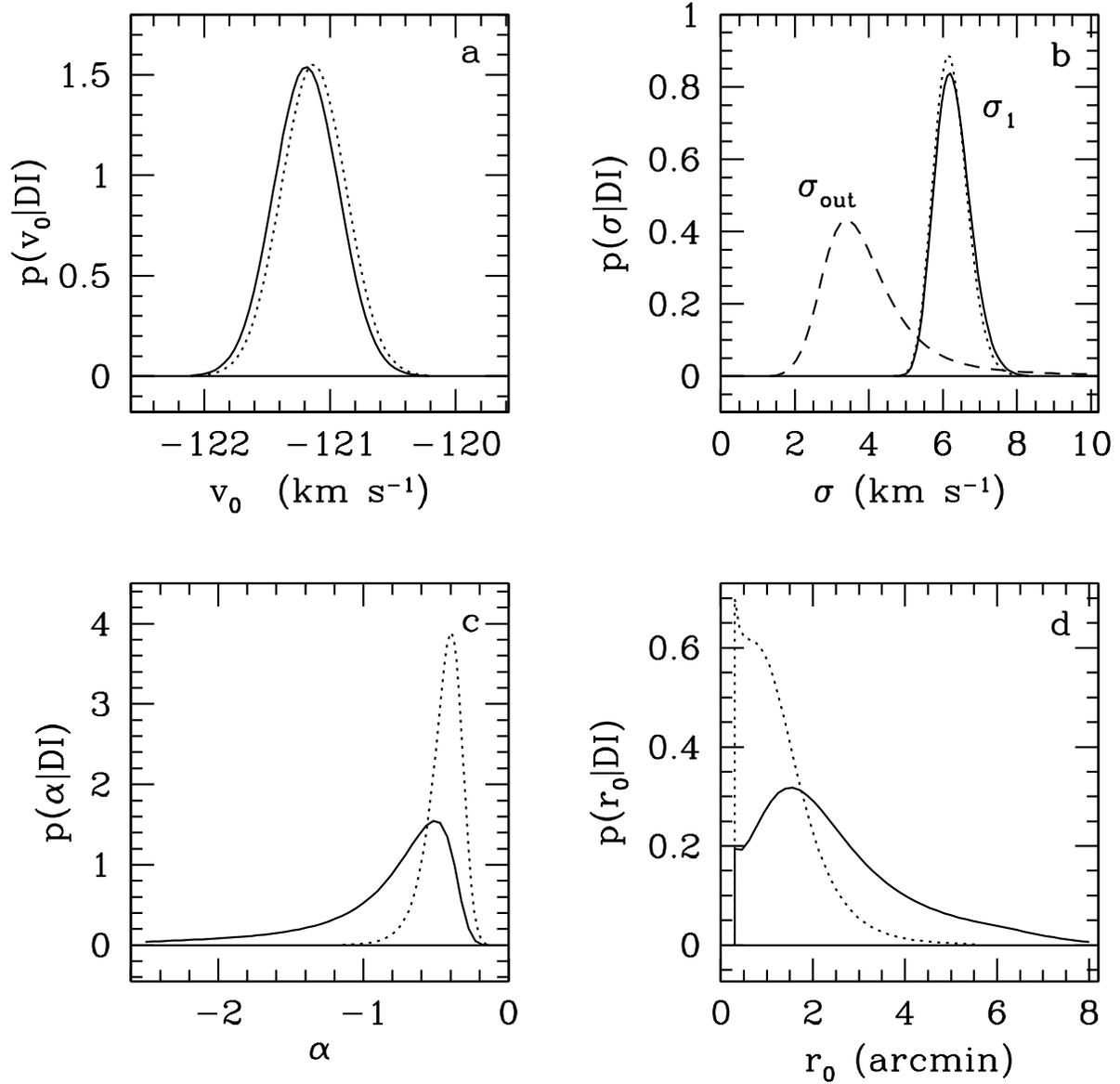} 
\figcaption[f10.eps]{Probability density
distributions for models $CB$ (solid) and $C$ (dotted) of the full M92
data set. (a) Mean velocity. (b) Velocity dispersion at 1\arcmin. The
dashed line shows the distribution for $\sigma_{\rm out}$ in the $CB$
model. The long tail to higher velocity is typical of dispersion
distribtutions based on only a few stars. (c) Power-law slope. (d)
Core radius.
\label{F:M92}}
\end{figure}

\begin{figure}
\plotone{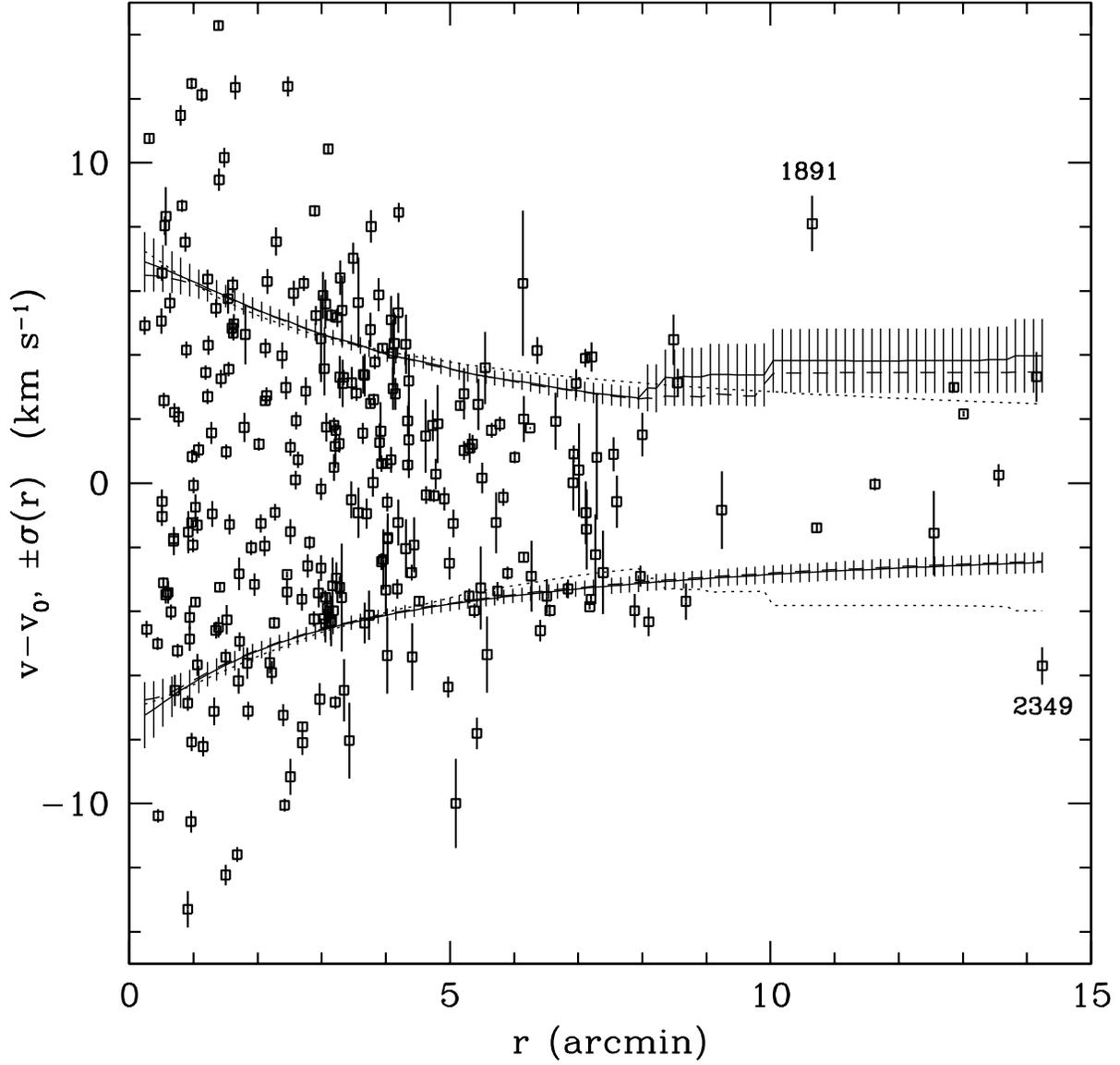} 
\figcaption[f11.eps]{Velocity disperion profiles
for our M92 data plotted with the velocities. The solid curves are the
mean of $p(\sigma|rDI)$ while the shading indicates the $1 \sigma$
errors based on this distribution. The dashed curves indicate the
modal results as discussed in the text. 
The upper curves
are for model $CB$, while the lower curves are for model $C$ . The mean curves for each distribution are
reflected about zero and plotted as dotted curves to ease
comparison. The two stars discussed in the text are labeled. 
\label{F:M92prof}}
\end{figure}

 The probability density distributions for the various parameters
in models $CB$ and $C$ are shown in Figure~\ref{F:M92}. The
probability distributions for the mean velocity and $\sigma_1$ are
consistent for both models and give
\momentsa{\bar v}{-121.2}{-121.2}{0.3} \kms\ and
\momentsa{\sigma_{1}}{6.3}{6.2}{0.5} \kms\ for model $C$ and
\momentsa{\bar v}{-121.1}{-121.1}{0.3} \kms\ and
\momentsa{\sigma_{1}}{6.2}{6.1}{0.5} \kms\ for model $CB$.
They do differ in the parameters giving the shape of the
velocity-dispersion profile. For model $CB$,
\moments{\alpha}{-0.8}{-0.5}{0.4} and
\moments{r_{0}}{2.5}{1.5}{1.6} arc minutes.
As can be seen in Figure~\ref{F:M92}, the probability distributions
for these quantities are broad and asymmetric. In addition, they are
anti-correlated; shallow profiles with narrow cores work equally
well as steep profiles with broad cores. 
The modal value for $r_{1}$ is at $10\arcmin$,
and it appears that the stars inside this radius can only weakly
constrain the shape of the curve. This can be seen graphically in 
Figure~\ref{F:M92prof}. (Fiducial points for the displayed M92
profiles are given in Table~\ref{T:M92prof}.)
The upper shaded region shows the 1-$\sigma$
spread in $p(\sigma|rD\hbox{CB}I_{\rm M92})$. Inside 8\arcmin, a
range of velocity profiles are possible. Outside this radius, the
profile is affected by the possible break radius and independent
velocity dispersion. (This can be seen in the step-like
increase in the outer part of the model $CB$. For such models, one of
the parameters
is the boundary $r_1$ and this parameter has a fairly wide range of 
values with non-trivial probability. Consequently this part of
the profile can be either inside or outside $r_{1}$ and the curve
reflects this ambiguity.) For these data and model $CB$,
\moments{\sigma_{\rm out}}{4.0}{3.4}{1.3} \kms. The asymmetry in the
probability distribution (see Figure~\ref{F:M92}b), is typical of cases where there
are few stars available to constrain the value of the dispersion.

For model $C$, \momentsa{\alpha}{-0.4}{-0.4}{0.1} and
\moments{r_{0}}{1.3}{0.3}{0.8} arc min. The former is well constrained,
but
the latter has its mode at the lower limit of its permitted range.
This indicates that there is some flattening to the profile, but the
core appears smaller than would be expected. The profile is shown 
as the lower shaded region in Figure~\ref{F:M92prof}. Model $C$ is 
strongly constrained
by the three outer stars with large deviations from the mean
velocity. In order to account for these velocities, a relatively
shallow, yet coreless profile is favored. Model $CB$ puts these
stars in the outer zone, removing their influence on the parameters
of the inner profile, but the velocities of the remaining stars can
be fit by a large range of profiles. Thus, model $CB$ is preferred.

The most deviant velocity is that of star 1891 in our list, lying at
$10\farcm 7$. Under some circumstances one might be tempted to
reject it outright from the cluster sample. This happens to be the
westernmost
star in our sample. It is fairly faint, but lies on the red giant
branch in the CMD. It has three relatively low quality velocity
measurements, but these  agree within their respective uncertainties.
\citet{tuc96} give the star only an 8\% probability of membership, but
this is not the only star with a low \citet{tuc96} probability that
satisfies our criteria for membership. A multi-component model such as
that described at the end of \S\ref{models} would directly incorporate
borderline cases such as this by assigning some probability to their
membership. For the present, we will just go to the other extreme and
redo the analysis without star 1891. 

The results for $\bar v$ and $\sigma_1$ are the same with or without
this star. For model $CB$, \moments{\sigma_{\rm out}}{3.9}{2.8}{1.7}
\kms,
the large skew and uncertainty reflecting the fact that this
distribution is 
derived from only the two outermost stars. For both models $CB$ and $C$, the probability distributions
for $\alpha$ and $r_0$ are somewhat broader and shifted with respect
to the distributions in Figure~\ref{F:M92}. For model $CB$
\moments{\alpha}{-0.9}{-0.6}{0.4} and \moments{r_{0}}{2.8}{1.8}{1.6}
arc min.
 As in the previous case,
$\sigma_1$ and $\alpha$ are strongly correlated with $r_0$. For model
$C$, \momentsa{\sigma_{1}}{6.3}{6.2}{0.5} \kms\ and
\momentsa{\alpha}{-0.5}{-0.5}{0.2}. There is now a peak to the
$r_{0}$ distribution and \moments{r_{0}}{1.5}{1.1}{0.9} arc min. 

From a comparison of the results of model $C$ probability
distributions for all possible subsamples obtainable by dropping a
single star, that obtained by dropping star 1891 stands out.  Dropping
this star gives a significantly larger increase in the model
probability than is the case for any other star. The only comparable
subsample is the one dropping  the outermost star (\#2349). In only
these two cases do the model parameters differ from those given by the
full sample. And it is only in these two cases that the mode of the
$r_{0}$ is not at the lower limit of the permitted range. Given these
results, it is possible that it is only these two stars that are
anomalous, rather than all the stars beyond a certain radius. Such a
suggestion constitutes a new model class, one where a certain fraction
of the stars are excluded, but we will not pursue this model here. We
will note, however, that a $C$-class model without either of these
stars gives \momentsa{\bar v}{121.1}{121.1}{0.2} \kms,
\momentsa{\sigma_{1}}{6.3}{6.2}{0.5} \kms,
\momentsa{\alpha}{-0.7}{-0.6}{0.3}, and
\moments{r_{0}}{2.2}{1.7}{1.2} arc min. 
From a consideration of the various samples and models discussed
above, it appears that $\sigma_{1}=6.3\pm 0.5$ \kms, $\alpha=-0.6\pm
0.3$, and $r_{0}=2\arcmin\pm 1\arcmin$ is a reasonable description of
the M92 velocity dispersion profile. For these values of $\alpha$ and
$r_{0}$ we can calculate $r_{c,v}=6\arcmin$, a value consistent with
our expectations based on the density profile as discussed in
\S\ref{S:Priors}.

There is little evidence from these data for a dominant population of
stars in the outskirts of M92 with higher-than-expected velocities. We
have, however, identified two stars which are highly inconsistent with
the velocity distribution of the remaining stars even though their
properties are otherwise consistent with cluster membership.  We will
discuss the possible origin of these stars following a reevaluation of
the M15 velocities.

\subsubsection{Rotation}

\begin{figure}
\plotone{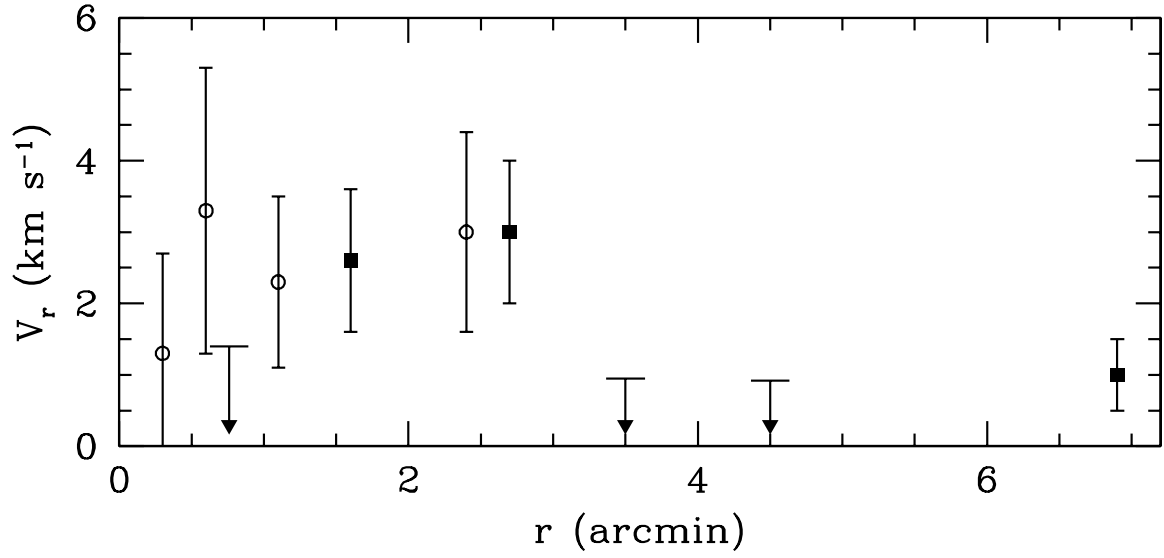}
\figcaption[f12.eps]{Rotation amplitude for M92 as a function of
radius. The first, fourth, and fifth points are $1 \sigma$ upper
limits. The open points are the bins with detected rotation from 
\protect{\cite{lup85}}.
\label{F:rotation}}
\end{figure}

Although this paper is focused on the velocity dispersion profile,
we would like to briefly follow up on the claim of
\citet{lup85} that the cluster is rotating. To that end, we've taken
our preferred M92 model from the previous section, assumed it to be
correct, and looked for sinusoidal rotational signals for six radial
bins. We then subtracted off this signal and estimated new
parameters for the profile model, iterating in this fashion until
the
values of the parameters stopped changing.  The parameters for 
the velocity dispersion profile are consistent with those given
above, except for $\sigma_{1}$ which drops to $6.0 \pm 0.5 \kms$.
The rotation profile is shown in Fig.~\ref{F:rotation} and is
compared to that of \citet{lup85}. (We've used their value of
$r_{c}=0.74$ pc and a distance of 8.2 kpc to convert their abscissae
to arc minutes.) A rotational
signal is only seen in half the bins; for the others a $1 \sigma$
upper limit is shown. The position angle is consistent for the bins
where a signal is seen.

\subsection{M15}
\begin{figure}
\plotone{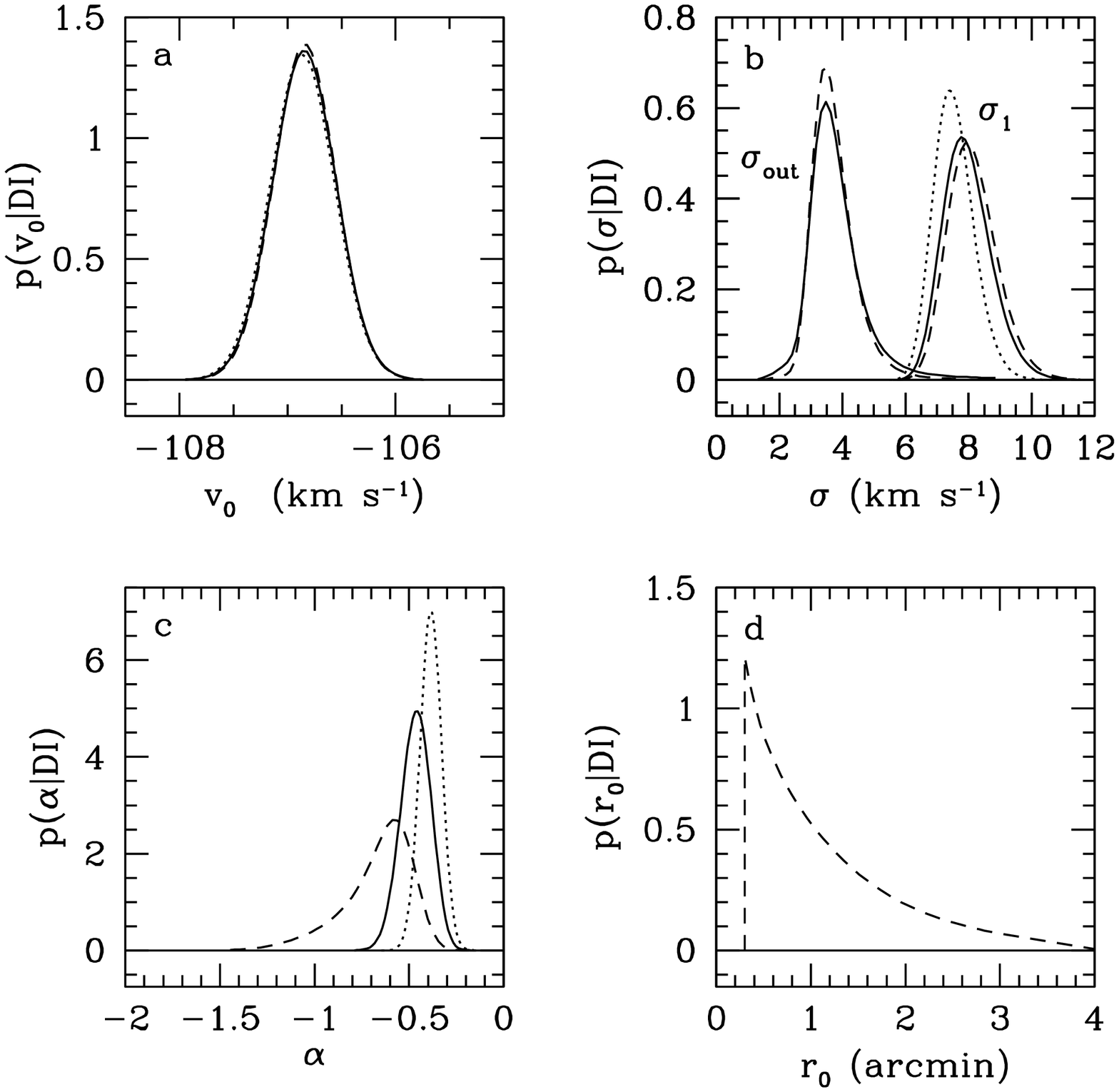}
\figcaption[f13.eps]{Probability density distributions for models $P$
(dotted), $PB$ (solid),
and $CB$ (dashed) of the M15 data set. (a) Mean velocity. (b) Velocity
dispersion at 1\arcmin ($\sigma_1$) and in the outer region
($\sigma_{\rm out}$). The outer region contains more stars than in the
comparable case in M92, so the distributions of $\sigma_{\rm out}$ are
less skewed.  (c) Power-law slope. (d) Core radius for model $CB$.
\label{F:M15}}
\end{figure}

In \citetalias{dru98} we concluded that there was an indication that
the velocity dispersion of M15 increased at large radii. In
Table~\ref{T:M15} we present our likelihoods and posterior
probabilities in the same way as in Table~\ref{T:M92}. In this case,
the most likely model is $CB$, but once our prior ratios are taken
into account, the most probable models are $P$ and  $PB$ by a small
margin. The probability densities are plotted in Figure~\ref{F:M15}.
None of the binned profiles is at all probable. 

\begin{figure}
\plotone{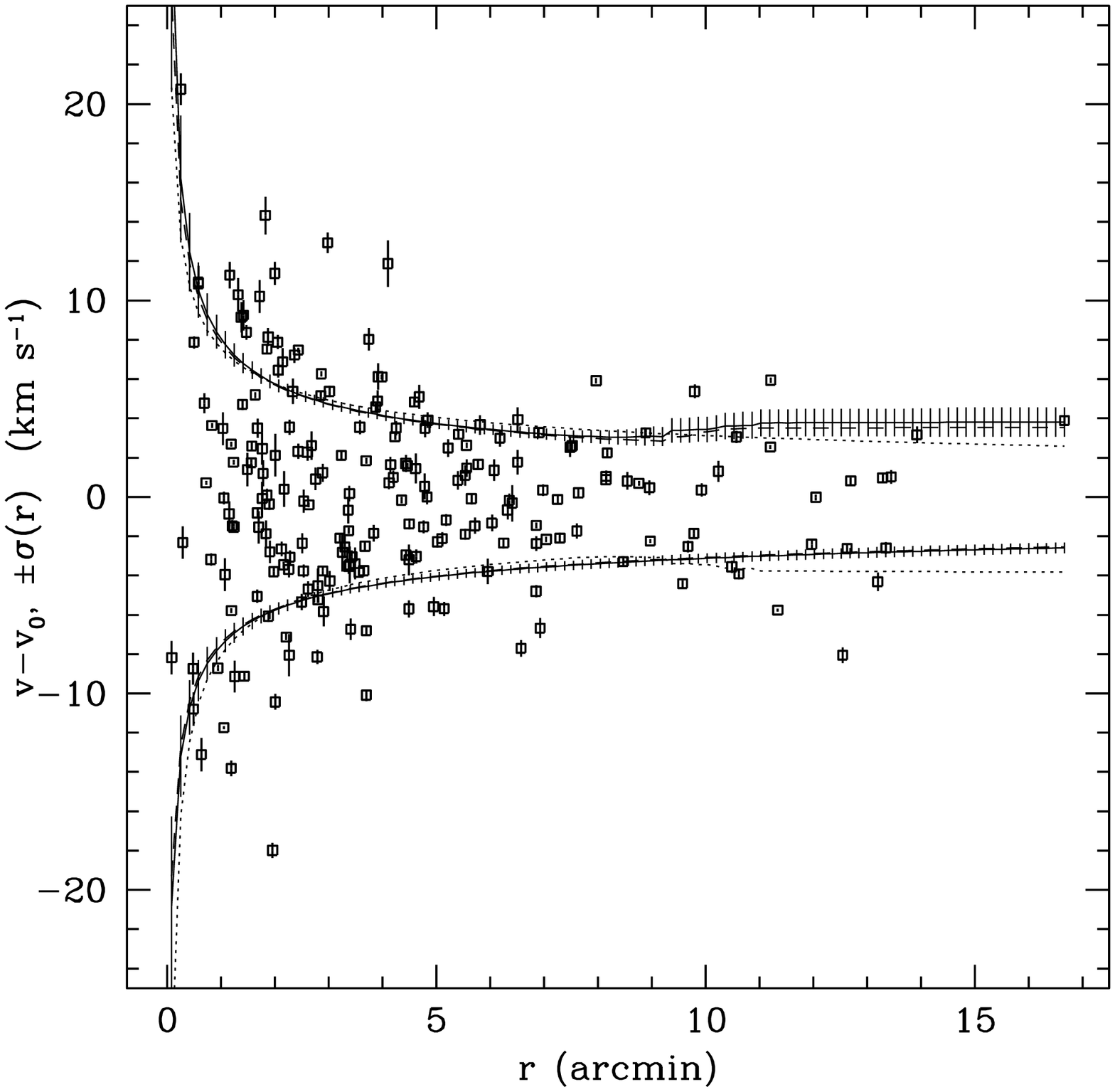} 
\figcaption[f14.eps]{As Figure~\ref{F:M92prof} 
for the M15 data from \citetalias{dru98}. The upper curve is for model
$PB$  and the lower for model $P$.  
\label{F:M15prof}}
\end{figure}

The higher likelihood of model $CB$ with respect to models $PB$ and
$P$ lies in its ability to give a slightly flatter profile
in the center.  In this case, the $r_0$ distribution is strongly
peaked at its lower limit, indicating that whatever core there is in
the velocity profile is unresolved. This is not unexpected, since as
we discussed in \S\ref{S:Priors}, the velocity dispersion profile in
M15 flattens at around 1\arcmin.  For model $CB$,
\momentsa{\sigma_{1}}{8.1}{8.0}{0.8} \kms, \momentsa{\alpha}{-0.7}{-0.6}{0.2},
\moments{r_{0}}{1.1}{0.3}{0.7} arc min, and \momentsa{\sigma_{\rm
out}}{3.7}{3.5}{0.7} \kms. Unlike M92, $\alpha$ and $r_0$ are
uncorrelated here.  The modal value for $r_{1}$ is 9\arcmin.

The profiles for models $P$ and $PB$ are shown 
in Figure~\ref{F:M15prof}.  The two models differ in the way they
accommodate the stars
beyond 9\arcmin. The slope and scale of the power-law are mainly set
by the high-velocity stars of the inner cluster.  For model $PB$
\momentsa{\sigma_{1}}{8.0}{7.8}{0.8} \kms,
\momentsa{\alpha}{-0.47}{-0.46}{0.08}, and \momentsa{\sigma_{\rm
out}}{3.8}{3.5}{0.9} \kms, with the modal value for $r_{1}$ at
9\arcmin. Model $P$ is constrained to a slightly shallower drop
(\momentsa{\alpha}{-0.39}{-0.39}{0.06}) and a smaller central amplitude
(\momentsa{\sigma_{1}}{7.5}{7.4}{0.6} \kms).  Nonetheless, the velocity
dispersion profile at large radii for model $P$ is less than that of
model $PB$. Model $PB$ fits the data somewhat better, but the
improvement is balanced by the need for two extra parameters and the
result is that the two descriptions remain equally
probable.  On the one hand, the data are compatible with a single power-law,
but on the other, they do not exclude at any level the possibility that 
some fraction of the stars outside about 9\arcmin\ have a somewhat higher
velocity dispersion than that expected by extrapolating the most probable power-law
for the stars inside that radius.

\section{Comparison with N-body  models}
 
In \citetalias{dru98} we suggested that the apparent flattening or upturn in the
velocity dispersion of M15 at large radii was due to the effect of the
Galactic tidal field. \citet{dru99}, however, showed that during core
collapse, a globular cluster ejects stars from the core. It is
therefore worthwhile to consider the effects of unbound stars
on the estimation of the velocity-dispersion profile. While the seemingly
systematic effect in M15 can be explained either way, the apparent high-velocity
members seen in M92 might more naturally be explained by the ejection mechanism.

    We have been carrying out a separate program to simulate the
       dynamic evolution of star clusters using GRAPE N-body
       supercomputers\footnote{See \citet{mak97} for a
       description of the GRAPE hardware development program.} at
       Indiana University.  We have run series of isolated cluster
       models with identical stars to well beyond core collapse, as a
       benchmark for comparison with simulations that include tidal
       effects.  Unlike our previous Fokker-Planck models of isolated
       clusters \citep{dru99}, full N-body models allow the
       possibility of stars acquiring positive energies,
       i.e. exceeding the escape velocity.  As noted by \citet{joh99}, the increasing dominance of the halo velocity
       distribution by escaping stars can produce a flattening of the
       halo velocity-dispersion profile.  We examine one such model
here to illustrate the effects on the velocity dispersion profile of
unbound stars expelled from the core.

       For comparison with the M92 results, we used N-body data from a
       GRAPE-4 run with N = 8192 identical point-mass stars.  The
       initial state was a Plummer model with no primordial binaries.
       We used the NBODY4 code \citep{aarseth99} to
evolve the model
       through core collapse and well into the post-collapse phase.
       Stellar escape occurred primarily as a result of cumulative
       energy increase from single-single scattering in the
       contracting core.  This produces an isotropic stream of
       escaping stars.  The escape rate increased as the model
       approached core collapse, with about 2.5\% of the cluster mass
       lost to escape at the time of core collapse.  As a compromise
between M92 and M15, we
selected a
       data snapshot from near the time of core collapse for analysis.

\begin{figure}
\plotone{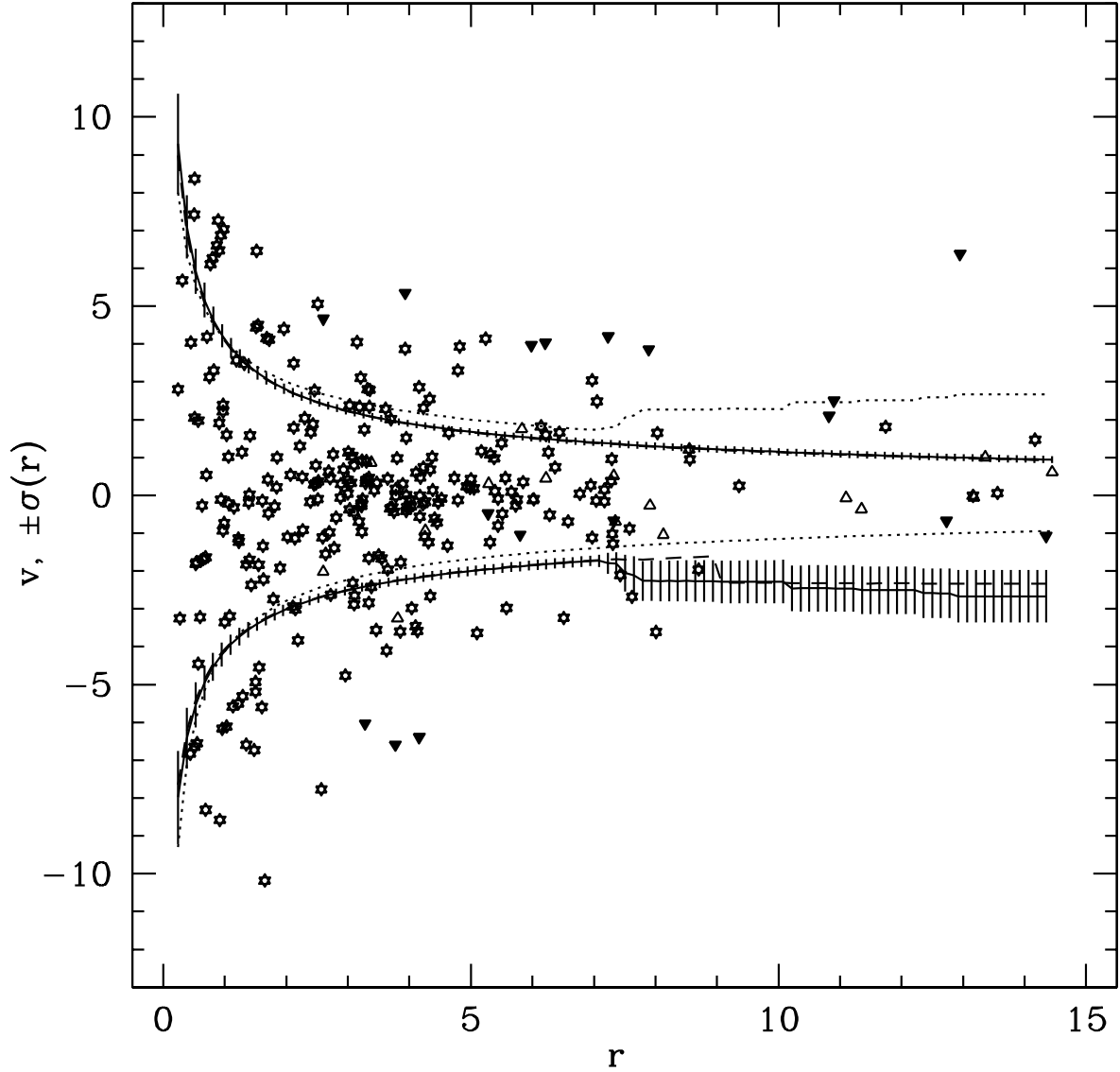}
\figcaption[f15.eps]{The stellar samples and most
probable profiles
for the $N$-body model. The stars indicate
stars selected for both the bound and unbound samples. The filled
triangles are unbound stars which replace the open triangles in the
unbound sample. The curves are as in Figure~\ref{F:M92prof}. The
upper shading is for model $P$ for the bound sample. The lower
shading is for model $PB$ for the unbound sample. 
\label{F:Xproj}}
\end{figure}

In order to compare this N-body model to the M92 data, we rescale the
N-body positions to give the model roughly the same limiting radius as
M92. The
velocities are scaled arbitrarily to give velocities with numerical
values similar to the M92 ones. 
We then selected stars from these projected lists to have roughly
the same radial distribution as the M92 data. We did this in two
ways. First we selected our sample from only the bound stars. Then
we  reselected the sample in the same way, but this time
allowed unbound stars to enter each sample. Each of these unbound
stars effectively replaced one bound star in the 
bound sample, so the total number of stars is maintained. In each
case a single, obvious, strongly unbound ``outlier'' has been
rejected. As in \S\ref{S:Examples}, the units for the parameters
should be obvious from the context. 

We show the results for a typical projection in
Figure~\ref{F:Xproj}. Stars common
to both samples are shown as hexagrams. Filled triangles are unbound
stars in the second sample. They replace the bound stars from the
first sample shown as open triangles. The shaded regions and curves
are as for the Figure~\ref{F:M92prof}. The best model class for the
bound
samples is shown at positive velocities, while that for the unbound
samples
is shown at negative velocities. The dotted lines show the reflections
of
the mean curves about zero velocity for the purposes of comparison. 

For the bound sample, 
models $P$ and $C$ are equally likely, but the $r_0$ distribution for
the latter is peaked towards the center, supporting the slightly
higher probability of model $P$. For model $P$,
\momentsa{\sigma_{1}}{4.1}{4.1}{0.3} and
\momentsa{\alpha}{-0.56}{-0.56}{0.06}.  We show model $P$. For the
unbound
sample, there are 17 unbound stars out of the remaining 279. The most
likely model is $B_{S=2}$ with
\momentsa{\sigma_{1}}{4.5}{4.4}{0.4}
 inside 1.76 and \momentsa{\sigma_{2}}{2.2}{2.2}{0.1} outside that
radius. A factor of three less likely, but more probable on physical
grounds, is model $PB$ (shown in the figure), with the outermost nine
stars in the outer zone for the modal value of $r_{1}=10$. Of these,
five are indeed unbound.  The outer dispersion is
\moments{\sigma_{\rm out}}{2.7}{2.3}{0.8}, but the difference between
the mean and the mode indicates a strong skew to higher values since
there are few data points to constrain it. In the inner region,
\momentsa{\sigma_{1}}{4.1}{4.1}{0.3} and
\momentsa{\alpha}{-0.44}{-0.44}{0.06}. The central scale is the same as
for the bound sample, but the slope of the power-law is now somewhat
flatter in order to take into account the velocities of the unbound
stars at intermediate radii. Model
$P$ is a factor of 9 less likely with \momentsa{\sigma_{1}}{3.9}{3.9}{0.3}
and
\momentsa{\alpha}{-0.36}{-0.36}{0.05}. The unbound star at a radius
of 12
affects the model in much the same way as star 1891 in M92. Removing
this star provides the greatest improvement to model $P$ and gives 
\momentsa{\sigma_{1}}{4.1}{4.0}{0.3} and
\momentsa{\alpha}{-0.43}{-0.43}{0.05}, nearly identical with the
previous model
$PB$  where this star is beyond $r_{1}$. The effect of the other
unbound stars remains, however. 

From this comparison, it appears that unbound stars can indeed modify
the inferred velocity dispersion profile, generally giving a shallower
slope than we find when only bound stars are used. Extreme outliers at
large radii can be identified by the effect their removal from the
sample has on the parameters and overall model probability.
On the basis of this admittedly limited examination it would appear
the two
rejected stars in M92 are likely to be unbound stars currently exiting
the cluster. 

 M15, being in deep core collapse, should have more escaping stars and
should show larger effects, but the details will depend on the radial
distribution of the unbound stars. It is possible that the sample
from \citetalias{dru98}
contains some unbound stars at relatively small radii and,
consequently, the
velocity dispersion profile has been inflated in the inner region.
The continuation to larger radii is consistent with the velocities
in that region, perhaps disguising the effects of high velocity
stars there. In any case, the ejection of  unbound stars from the
cluster core is a stochastic
process and not all snapshots or projections will contain such
stars, or have the same radial distribution of unbound stars. 
What this example does demonstrate, however, is that  sporadic star
ejection from the core can produce an apparent increase in the
velocity dispersion in the outer region. This is especially so
considering that, going out from the core, the unbound
stars form an increasing fraction of
 the cluster-associated stars.

\section{Summary}
We have carried out an extensive investigation of the global radial
velocity field in M92, using the WIYN telescope for both photometry
and spectroscopy.  We obtained high-accuracy (median error 0.35 \kms) velocities
for 299 probable cluster members, thereby greatly increasing the
number of stars with measured radial velocities in this rich cluster,
which has a very high density, yet resolved core.  This new data set
nicely complements that from our previous study of the collapsed-core
cluster M15 \citepalias{dru98}.

We selected likely cluster candidate members for spectroscopy by
several photometric methods.  The most efficient of these involved
obtaining a large-area, 3-band Washington photometry mosaic of M92
with WIYN\@.  Using a photometric metallicity index, we were able to
successfully identify cluster members with $\sim$ 70\% efficiency
within the projected cluster halo ($r>3'$), where the vast majority of
stars are nonmembers.

We've introduced an improved Bayesian analysis scheme and have applied
 it to both this data set and the M15 data set of \citetalias{dru98}.
 Of our models, the most probable one for the M92 data is one
 described by a core radius of about 2\arcmin, velocity dispersion at
1\arcmin\ of 6.3 \kms, and an outer power-law slope of between $-0.5$
and 
 $-0.8$ depending on which, if any, of two, outer, high-velocity stars
are included. It is these stars that appear to increase the
dispersion in the outer part of the cluster.
 The M15
 velocities can be described equally well by both a single power-law
with slope of $-0.5$ and velocity dispersion at 1\arcmin\ of 8.0 \kms,
or by an inner power-law with slope $-0.4$ and scale of 7.5 \kms\ plus
an outer region beyond 9\arcmin\ with a dispersion of 4 \kms. This
region could be populated by some fraction of unbound stars. It is not
unlikely that a few unbound stars are also present in the inner region
of the cluster, particularly between 2\arcmin\ and 5\arcmin. 
There is also some evidence that the slope of the power-law flattens
at the inner limit of our sample. 

Our consideration of the velocity profiles in a GRAPE-based N-body
model of an isolated cluster suggests that our two data sets may
contain a number of escaping stars that were boosted to positive
energies by the same internal relaxation processes that drive core
collapse.  Clearly, more data are needed to answer the questions of
whether the velocity profiles do indeed flatten at large radii and
to identify the physical mechanism behind this. 
For a realistic cluster in a
tidal environment,
the development of the halo must involve an interplay between internal
two-body relaxation and external tidal influences.  

Measures of mass segregation and velocity anisotropy may provide the
best means to gauge the relative importance of two-body relaxation and
tidal effects in determining the structure of cluster halos.  Mass
segregation is best investigated by very deep imaging.  \citet{and00}
have used Hubble Space Telescope WFPC2 photometry to investigate mass
segregation in M92\@.  They find a significant change in the slope of
the mass function between fields at 11 and 19 core radii (2\farcm5 and
4\farcm3, respectively), in the sense of a steepening with increasing
radius as expected from two-body relaxation.  It would be useful to
determine the mass function, in a similar way, at an even greater
distance from the cluster center in order to determine whether this
steepening continues into the outer halo.  Given the sharp drop of
stellar surface density in the halo, this may require larger area,
ultra-deep, ground-based imaging to obtain sufficient statistics.

Radial velocities, alone, do not strongly constrain the anisotropy
profile.  Both radially-biased and
tangentially-biased orbit distributions are capable of producing a
halo flattening of the projected velocity-dispersion profile.  In
contrast, the anisotropy profile can be directly determined from
proper motions, either alone \citep[e.g.][for M3]{cud79} or in
conjunction with radial velocities \citep[e.g.][for M13]{lup87}.  In
the case of M92, \citet{cud79} finds little evidence for anisotropy,
although it appears that this data set would only be capable of
detecting a strong signal. 

\citet{tes00} have used the Digitized Palomar Observatory Sky Survey
(DPOSS) to examine the spatial structure of M92 near and beyond the
$\sim 15\arcmin$ tidal radius.  They found clear evidence for a
flattening of the profile beyond their best fit tidal radius of $r_t =
12\farcm3$, which they interpret as a extra-tidal halo that extends to
approximately $30\arcmin$.  The structure of this halo is essentially
circular, showing at best weak evidence for the sort of bipolar
elongation---``tails''---seen in some other clusters \citep[e.g.\
Palomar 5;][]{ode01}.  \cite{tes00} suggest that this may indicate
that the extra-tidal halo of M92 consists of stars evaporating from
the cluster, which have not yet formed an escaping tidal stream.  This
would tend to argue that two-body relaxation and possibly tidal
shocking dominate over tidal stresses in determining the halo
structure and mass loss rate rate in M92\@.  

Improved dynamical models for clusters, which take advantage of the
growing datasets provided by high resolution imaging and spectroscopy,
should provide a more sensitive measure of the role of tidal
influences on cluster halos.  This will lead to improved estimates of
cluster mass loss rates and dissolution timescales.

\acknowledgments
We'd like to thank Doug Geisler for his advice on using the
Washington system and Heather Morrison for making her 
data available to us. G. A. D. was supported in part by NASA LTSA
grant NAG 5-6404 and, during the completion of this study, by grants
for the following two HST projects. Support for Programs numbered GO-9839
and AR-10292 was provided by NASA through a grant from the Space
Telescope Science Institute, which is operated by the Association of
Universities for Research in Astronomy, Incorporated, under NASA
contract NAS5-26555. The GRAPE simulations were supported by NASA ATP
grant NAG 5-2781 to Indiana University and also received generous
support from Indiana University Information Technology
 Services. S. S. acknowledges support of the School of Engineering,
Mathematics and Science at Purdue University Calumet for publication
costs.

\clearpage

\begin{deluxetable}{lcc}
\tablecaption{Comparison of M92 and M15. \label{T:M92-M15}}
\tablewidth{0pt}
\tablehead{
\colhead{Property\tablenotemark{a}} & \colhead{M92}   & \colhead{M15} 
}
\startdata
$M_V$ & $-$8.20 & $-$9.17 \\
$\log \rho_0 ~(L_\sun\,\pc^{-3})$ & 4.29 & $>\,$5.38 \\
$r_c$ ~(\arcmin) & 0.23 & $<\,$0.025\tablenotemark{b} \\
$r_h$ ~(\arcmin)                        & 1.09          & 1.06 \\
$r_t$ ~(\arcmin) & 15.2 & 21.5      \\
$c$ & 1.81 & $>\,$2.5  \\
\mbox{[Fe/H]} & $-$2.29 & $-$2.25 \\
$R_{\sun}$ ~(kpc) & 8.2 & 10.3    \\
$R_{\rm GC}$ ~(kpc) & 9.6 & 10.4    \\
$Z$ ~(kpc) & 4.7 & $-$4.7  \\
\enddata

\tablenotetext{a}{These data are from the online compilation of
\citet{har96}.}
\tablenotetext{b}{\citet{sos97}}
\tablecomments{\\
$M_V$ = total absolute magnitude \\
$\rho_0$ = central luminosity density \\
$r_c$ = core radius \\
$r_h$ = half-mass radius \\
$r_t$ = tidal radius \\
$c$ = concentration parameter \\
$[$Fe/H$]$ = metallicity \\
$R_\sun$ = solar distance \\
$R_{\rm GC}$ = galactocentric distance \\
$Z$ = distance from galactic plane
}
\end{deluxetable}

\begin{deluxetable}{lcccc}
\tablecaption{Observing Log for M92\label{T:Obs}}
\tablewidth{0pt}
\tablehead{\colhead{Configuration} &\colhead{UT Date} & \colhead{HJD} & \colhead{Exposure time} &
\colhead{\#}}
\startdata
m1  &  1996 May 14 & 10217.799  & 2400s & 38  \\
m2  &  1996 May 16 & 10219.757 & 1800s & 26  \\
m3  &  1996 May 17 & 10220.764 & 1800s & 27  \\
m4  &  1996 May 17 & 10220.852 & 3000s & 24   \\
m5  &  1996 May 18 & 10221.769 & 3000s & 32   \\
m6  &  1996 May 19 & 10222.743 & 4200s\tablenotemark{a} &  29\\
m7  &  1996 May 19 & 10222.845 & 3600s &  31   \\
M6  &  1996 June 19 & 10253.685 & 1800s & 29   \\
M7  &  1996 June 19 & 10253.740 & 1800s & 32 \\
M8  &  1996 June 19 & 10253.785 & 1800s & 31 \\
J1  &  1996 June 21 & 10255.687 & 1800s & 88 \\
J2  &  1996 June 21 & 10255.747 & 1800s & 87 \\
J3a &  1996 June 21 & 10255.818 & 991s\tablenotemark{a} & 85 \\
J3b &  1996 June 23 & 10256.724 & 1800s  & 85 \\
J4  &  1996 June 23 & 10256.793 & 1800s & 77 \\
J5  &  1996 June 23 & 10257.803 & 1800s & 66 \\
A   &  1997 June 22 & 10621.724 & 7200s & 71 \\
B   &  1997 June 22 & 10621.875 & 7200s& 69 \\
C   &  1997 June 23 & 10622.727 & 7200s& 60 \\
D   &  1997 June 23 & 10622.888 & 6760s& 57 \\
E   &  1997 June 24 & 10623.708 & 7200s& 57 \\
F   &  1997 June 24 & 10623.833 & 7200s& 53 \\
G   &  1997 June 25 & 10624.723 & 7200s& 54 \\
H   &  1997 June 25 & 10624.882 & 7200s& 58 \\
K   &  1997 June 26 & 10625.725 & 7200s& 55 \\
L   &  1997 June 26 & 10621.853 & 7200s& 54 \\
M   &  1998 June 16 & 10980.690 & 7200s& 69 \\
\enddata
\tablenotetext{a}{Observation affected by clouds}
\end{deluxetable}

\begin{deluxetable}{llrrrrrrrrl}
\tablecaption{Data for cluster members \label{T:members}}
\tablewidth{0pt}
\rotate
\tablehead{\colhead{R.A.} & \colhead{Decl.} & & \colhead{r}& & & &  & \colhead{$v$} & \colhead{$\epsilon_v$} & \colhead{Notes\tablenotemark{b}  and} \\
\colhead{(J2000.0)} & \colhead{(J2000.0)} & \colhead{ID} & \colhead{(\arcmin)} & \colhead{T2} & \colhead{M-T2} & \colhead{C-M} & \colhead{$N_v$\tablenotemark{a}} & \colhead{(\kms)} & \colhead{(\kms)}   & \colhead{Other Names}\\
\colhead{(1)} &
\colhead{(2)} &
\colhead{(3)} &
\colhead{(4)} &
\colhead{(5)} &
\colhead{(6)} &
\colhead{(7)} &
\colhead{(8)} &
\colhead{(9)} &
\colhead{(10)} &
\colhead{(11)} 
}
\startdata
17 16 19.65 & +43 14 26.0 &     1891 & 10.65 & 15.44 &  1.03 &  0.56 &   3 & -113.08 & 0.87 & T \\
17 16 22.24 & +42 58 04.9 &      516 & 13.01 & 13.62 &  1.09 &  0.75 &  13 & -119.02 & 0.09 & T \\
17 16 24.16 & +43 07 12.4 &    VII-3 &  7.88 & 15.45 &  1.06 &  0.53 &   2 & -125.16 & 0.52 & T \\
17 16 24.38 & +43 11 45.5 &     1924 &  8.55 & 15.47 &  1.02 &  0.49 &   4 & -118.05 & 0.44 & T \\
17 16 27.68 & +43 08 55.1 &     2343 &  7.21 & 15.79 &  1.02 &  0.54 &   3 & -117.24 & 0.45 &  \\
17 16 29.83 & +43 06 04.3 &    VII-8 &  7.11 & 14.94 &  1.10 &  0.58 &   4 & -117.28 & 0.29 & T \\
17 16 29.99 & +43 02 45.3 &     1995 &  8.68 & 15.53 &  0.96 &  0.55 &   4 & -124.87 & 0.57 & T \\
17 16 31.13 & +43 04 46.5 &     3917 &  7.39 & 16.77 &  0.93 &  0.47 &   2 & -123.97 & 1.30 &  \\
17 16 33.36 & +43 10 21.8 &     1936 &  6.51 & 15.48 &  1.03 &  0.54 &   3 & -124.71 & 0.54 & T \\
17 16 33.89 & +43 07 04.0 &     3180 &  6.15 & 16.37 &  0.96 &  0.50 &   3 & -119.18 & 0.72 &  \\
\enddata
\tablecomments{The full table is available on-line as a machine-readable table.}
\tablenotetext{a}{Number of observations. Where there is only one observation,
the configuration code from Table~\ref{T:Obs} is given. For stars with configuration 'Z',
the velocity is based on the sum of all available spectra.}
\tablenotetext{b}{R: Has a proper motion in \citet{ree92}. T: Has a proper motion in \citet{tuc96}. v: Variable in our data. V: Known to be variable from another source. j: Velocities likely affected by jitter. p: Photometry is from a lower-quality source.}
\end{deluxetable}

\begin{deluxetable}{lllrrrrrrrl}
\tablecaption{Data for doubtful cluster members \label{T:doubt}}
\tablewidth{0pt}
\rotate
\tablehead{\colhead{R.A.} & \colhead{Decl.} & & \colhead{r}& & & &  & \colhead{$v$} & \colhead{$\epsilon_v$} &\colhead{Notes\tablenotemark{b}  and} \\
\colhead{(J2000.0)} & \colhead{(J2000.0)} & \colhead{ID} & \colhead{(\arcmin)} & \colhead{T2} & \colhead{M-T2} & \colhead{C-M} & \colhead{$N_v$\tablenotemark{a}} & \colhead{(\kms)} & \colhead{(\kms)}   & \colhead{Other Names}\\
\colhead{(1)} &
\colhead{(2)} &
\colhead{(3)} &
\colhead{(4)} &
\colhead{(5)} &
\colhead{(6)} &
\colhead{(7)} &
\colhead{(8)} &
\colhead{(9)} &
\colhead{(10)} &
\colhead{(11)} 
}
\startdata
17 15 59.02 & +43 09 53.3 &     2171 & 12.52 & 15.67 &  0.88 &  0.55 &   5 &  -89.33 &
0.39 & To \\
17 16 19.64 & +43 14 03.8 &     2919 & 10.44 & 16.20 &  0.91 &  0.49 &   3 & -100.89 & 1.06 &
To \\
17 17 32.80 & +43 12 00.4 &     3448 &  6.05 & 16.52 &  1.07 &  0.62 &   3 &  -94.31 & 0.88 &
o \\
17 18 12.18 & +42 55 04.0 &       30 & 17.74 & 10.26 &  2.03 &  1.70 &   1 & -108.45 & 1.12 &
Tpo \\
\\
17 16 29.11 & +43 09 44.9 &     VI-7 &  7.09 & 12.66 &  1.03 &  0.65 &  m1 & -139.57 & 0.31 & RTc ZNG-4 \\
\\
17 17 07.91 & +43 07 12.2 &     R644 &  1.00 & 14.03 &  1.09 &  1.06 &   3 & -121.84 & 0.92 & Rh \\
17 18 39.80 & +43 06 10.5 &     1016 & 17.05 & 14.91 &  0.70 &  0.70 &  J2 &  -85.97 & 0.87 & Th \\
\enddata
\tablenotetext{a}{Number of observations. Where there is only one observation,
the configuration code from Table~\ref{T:Obs} is given.}
\tablenotetext{b}{R: Has a proper motion in \citet{ree92}. T: Has a proper motion in \citet{tuc96}.
o: Outlying velocity c: Odd colors. h: High $h_1$
value. p: Photometry is from a lower-quality source.}
\end{deluxetable}

\begin{deluxetable}{lllrrrrrrl}
\tablecaption{Data for non-members \label{T:non}}
\tablewidth{0pt}
\rotate
\tablehead{\colhead{R.A.} & 
\colhead{Decl.} & 
& 
& 
& 
&  
& \colhead{$v$} 
& \colhead{$\epsilon_v$} &
\colhead{Notes\tablenotemark{b}  and }\\
\colhead{(J2000.0)} &
\colhead{(J2000.0)} &
\colhead{ID} &
\colhead{T2} &
\colhead{M-T2} & 
\colhead{C-M} & 
\colhead{$N_v$\tablenotemark{a}} & 
\colhead{(\kms)} & 
\colhead{(\kms)}  &  
\colhead{Other Names}\\
\colhead{(1)} &
\colhead{(2)} &
\colhead{(3)} &
\colhead{(4)} &
\colhead{(5)} &
\colhead{(6)} &
\colhead{(7)} &
\colhead{(8)} &
\colhead{(9)} &
\colhead{(10)} 
}
\startdata
17 14 16.43 & +43 09 40.8 &       40 & 10.86 &  1.42 & \nodata & J3a &  -65.95 & 2.35 & p \\
17 14 24.38 & +42 50 39.0 &      594 & 13.42 &  1.36 & \nodata &   2 &  -89.90 & 0.44 & p \\
17 14 24.44 & +42 55 45.8 &      487 & 13.70 &  0.90 &  0.47 & J3a &  -47.57 & 2.27 & p \\
17 14 27.31 & +42 44 50.4 &      389 & 13.11 &  1.24 & \nodata &  J4 &   -3.30 & 1.53 & p \\
17 14 30.21 & +43 03 26.5 &      447 & 13.43 &  0.95 &  0.68 &  J1 &   -2.41 & 0.76 & p \\
17 14 32.91 & +43 01 10.6 &      739 & 13.90 &  1.20 &  0.90 &   2 &  -95.67 & 0.96 & p \\
\enddata
\tablecomments{The full table is available on-line as a machine-readable table.}
\tablenotetext{a}{Number of observations. Where there is only one observation,
the configuration code from Table~\ref{T:Obs} is given. For stars with configuration 'Z',
the velocity is based on the sum of all available spectra.}
\tablenotetext{b}{R: Has a proper motion in \citet{ree92} and probablity of
 membership $\ge 90\%$. r: Has a proper motion in \citet{ree92} and probablity of 
membership $< 90\%$. T: Has a proper motion in \citet{tuc96} and probablity of
 membership $\ge 90\%$.t: Has a proper motion in \citet{tuc96} and probablity of
 membership $< 90\%$. v: Variable. p: Photometry is from a lower-quality source.}
\tablenotetext{c}{This identification is ambiguous. This star could be R137.}
\end{deluxetable}

\clearpage

\begin{deluxetable}{llcccc}
\tablecaption{Permitted ranges for parameters \label{T:range}}
\tablewidth{0pt}
\tablehead {
                  &  & \multicolumn{2}{c}{M92} &
\multicolumn{2}{c}{M15} \\
\colhead{Parameter}& \colhead{Models} & \colhead{lower} &
\colhead{upper} &
\colhead{lower} & \colhead{upper} }
\startdata
$\bar v$ (\kms)      & All &  -122.58     &   -119.58 &  -108.5 &  -105 \\
$\sigma_{1}$ (\kms)   & All &   0.3        &   12      &   0.3   & 20
\\
$\alpha $         & $P$,$PB$,$C$,$CB$ &  -2.5       &   0       &  -2.5
 
& 0
\\
$r_0$ (\arcmin)   & $C$,$CB$ & 0.3         &   8       &   0.3   & 3 \\
$r_1$ (\arcmin)   & $PB$,$CB$ & 8           &   14      &   7.5   & 15
\\
$\sigma_{\rm out}$ (\kms) & $PB$,$CB$ & 0.3  &    10      &   0.3   & 10 \\
\enddata
\end{deluxetable}
\clearpage

\begin{deluxetable}{lccccc}
\tablecaption{Results for Mock Data Sets \label{T:mock}}
\tablewidth{0pt}
\tablehead{
               & \multicolumn{5}{c}{$\log(p(\hbox{Model}| D I)/p(\hbox{Best Model}|DI))$}\\
\colhead{Model} & \colhead{\bf PL} & \colhead{\bf CPL} &
\colhead{\bf PL+F}&
\colhead{\bf D}&
\colhead{\bf S}
}
\startdata
$P$             &   0.0        &  -1.2      &   -7.3    &   -5.8      &    -2.1    \\
$C$             &  -1.7       &   0.0       &   -8.8    &   -4.0      &    -1.8    \\
$PB$            &  -0.5       &   -1.7      &   -0.04   &   -6.6      &    -3.7    \\
$CB$            &  -2.3       &   -0.3      &    0.0    &   -5.1       &    -3.4    \\
$B_N$ (1-20)     &  -9.5       &   -3.0     &   -5.7    &   -1.5      &    0.1     \\
$B_N$ (best)     &  -9.9 (11)  &   -3.3 (7) &   -6.1 (7)&   -1.5 (5)  &    0.0 (1)   \\
$B_S$ (2-4)      &  -10.5      &   -4.3     &   -6.9    &    0.1      &    0.0      \\
$B_S$ (best)     &  -10.5 (4)  &   -4.3 (3) &   -6.9 (4)&    0.0 (3)  &   -0.1 (2) \\
\enddata
\end{deluxetable}
\clearpage

\begin{deluxetable}{lccc}
\tablecaption{Results for M92 \label{T:M92}}
\tablewidth{0pt}
\tablehead{
 & \colhead{$\log\left[{p(D|H I)\over p(D|CB\
I)}\right]$}
                & \colhead{$\log\left[{p(H| I)\over p(CB|I)}\right]$}
                & \colhead{$\log\left[{p(H|D I)\over
p(CB|DI)}\right]$}\\
\colhead{Model}& \colhead{(1)} &
\colhead{(2)} &
\colhead{(3)} 
}
\startdata
$P$             &  -0.5       &  -2.0  &   -2.5  \\
$C$             &  -0.2       &   0.0  &   -0.2  \\
$PB$            &  -0.9       &  -2.0 &   -2.9   \\
$CB$            &   0.0       &   0.0 &    0.0   \\
$B$             &   0.1      & -1.7 &  -1.6      \\
\hline
\phs$B_{N=2}$       &   0.0       &  -2.9 &   -2.9    \\
\phs$B_{N=3}$       &  -1.4       &  -2.9  &   -4.3   \\
\phs$B_{N=4}$       &  -1.5       &  -2.9  &   -4.4   \\
\phs$B_{N=5}$       &  -2.1       &  -2.9  &   -5.0   \\
\phs$B_{N=6}$       &  -3.2       &  -2.9  &   -6.1   \\
\phs$B_{S=2}$       &  -0.7       &  -2.9  &   -3.6   \\
\phs$B_{S=3}$       &  -1.1       &  -2.9  &   -4.0   \\
\phs$B_{S=4}$       &  -1.8       &  -2.9  &   -4.7   \\
\enddata
\tablecomments{(1) Log relative likelihood,
(2) Log relative prior
probability,
(3) Log relative posterior probability}
\end{deluxetable}

\begin{deluxetable}{lrrrr}
\tablecaption{Mean points for M92 \label{T:M92prof}}
\tablewidth{0pt}
\tablehead{
 & \multicolumn{2}{c}{Model C} & \multicolumn{2}{c}{Model CB}\\
\colhead{r} & \colhead{ $\langle\sigma(r)\rangle$}
& \colhead{$\epsilon_{\sigma(r)}$} &
\colhead{ $\langle\sigma(r)\rangle$} & \colhead{$\epsilon_{\sigma(r)}$}\\
\colhead{(\arcmin)} &
\colhead{(\kms)}&\colhead{(\kms)}&
\colhead{(\kms)}&\colhead{(\kms)}}

\startdata
 0.24 &  7.25 & 1.06 &  6.90 & 0.93 \\
 0.80 &  6.54 & 0.57 &  6.46 & 0.56 \\
 1.36 &  5.89 & 0.36 &  5.96 & 0.38 \\
 1.92 &  5.35 & 0.29 &  5.48 & 0.31 \\
 2.48 &  4.90 & 0.26 &  5.04 & 0.28 \\
 3.04 &  4.54 & 0.24 &  4.64 & 0.26 \\
 3.60 &  4.24 & 0.22 &  4.31 & 0.28 \\
 4.16 &  3.98 & 0.22 &  3.95 & 0.22 \\
 4.72 &  3.76 & 0.21 &  3.73 & 0.23 \\
 5.28 &  3.58 & 0.22 &  3.44 & 0.26 \\
 5.84 &  3.41 & 0.22 &  3.25 & 0.24 \\
 6.40 &  3.27 & 0.23 &  3.07 & 0.28 \\
 6.96 &  3.14 & 0.24 &  2.89 & 0.32 \\
 7.52 &  3.03 & 0.25 &  2.75 & 0.33 \\
 8.08 &  2.93 & 0.26 &  2.97 & 0.72 \\
 8.64 &  2.84 & 0.26 &  3.32 & 0.89 \\
 9.20 &  2.75 & 0.27 &  3.39 & 0.94 \\
 9.76 &  2.68 & 0.28 &  3.37 & 0.97 \\
10.32 &  2.61 & 0.29 &  3.82 & 0.98 \\
10.88 &  2.54 & 0.29 &  3.81 & 0.98 \\
11.44 &  2.48 & 0.30 &  3.81 & 0.99 \\
12.00 &  2.43 & 0.31 &  3.81 & 0.99 \\
12.56 &  2.38 & 0.31 &  3.81 & 0.99 \\
13.12 &  2.33 & 0.32 &  3.82 & 1.00 \\
13.68 &  2.28 & 0.32 &  3.84 & 1.02 \\
14.24 &  2.24 & 0.32 &  3.97 & 1.14 \\
\enddata
\end{deluxetable}

\begin{deluxetable}{lccc}
\tablecaption{Results for M15 \label{T:M15}}
\tablewidth{0pt}
\tablehead{
& \colhead{$\log\left[{p(D|H I)\over p(D|CB\
I)}\right]$}
                & \colhead{$\log\left[{p(H| I)\over p(CB|I)}\right]$}
                & \colhead{$\log\left[{p(H|D I)\over
p(CB|DI)}\right]$}\\
\colhead{Model}& \colhead{(1)} &
\colhead{(2)} &
\colhead{(3)} 
}
\startdata
$P$             &  -0.3       &   0.0  &   -0.3   \\
$C$             &  -0.6       &  -0.7  &   -1.3   \\
$PB$            &  -0.3       &   0.0 &   -0.3   \\
$CB$            &   0.0       &  -0.7 &   -0.7   \\
$B$             &   -0.9    & -1.7 & -2.6 \\
\hline
\phs$B_{N=2}$       &  -3.2       &  -2.9  &   -6.1    \\
\phs$B_{N=3}$       &  -0.9       &  -2.9  &   -3.8    \\
\phs$B_{N=5}$       &  -3.0       &  -2.9  &   -5.9    \\
\phs$B_{N=6}$       &  -3.4       &  -2.9  &   -6.3    \\
\phs$B_{N=7}$       &  -2.8       &  -2.9  &   -5.7    \\
\phs$B_{S=2}$       &  -1.9       &  -2.9  &   -4.8     \\
\phs$B_{S=3}$       &  -1.7       &  -2.9  &   -4.6     \\
\phs$B_{S=4}$       &  -1.9       &  -2.9  &   -4.8     \\
\enddata
\tablecomments{(1) Log relative likelihood,
(2) Log relative prior
probability,
(3) Log relative posterior probability}
\end{deluxetable}

\begin{deluxetable}{lrrrr}
\tablecaption{Mean points for M15 \label{T:M15prof}}
\tablewidth{0pt}
\tablehead{
 & \multicolumn{2}{c}{Model P} & \multicolumn{2}{c}{Model PB}\\
\colhead{r} & \colhead{ $\langle\sigma(r)\rangle$}
& \colhead{$\epsilon_{\sigma(r)}$} &
\colhead{ $\langle\sigma(r)\rangle$} & \colhead{$\epsilon_{\sigma(r)}$}\\
\colhead{(\arcmin)} &
\colhead{(\kms)}&\colhead{(\kms)}&
\colhead{(\kms)}&\colhead{(\kms)}}

\startdata
 0.08 & 20.85  & 4.60 & 28.99 & 8.38 \\
 0.74 &  8.47 & 0.83 &  9.29  & 1.08 \\
 1.41 &  6.59 & 0.46 &  6.80  & 0.51 \\
 2.07 &  5.68 & 0.32 &  5.65  & 0.33 \\
 2.73 &  5.10 & 0.26 &  4.96  & 0.26 \\
 3.40 &  4.69 & 0.23 &  4.47  & 0.25 \\
 4.06 &  4.38 & 0.22 &  4.12  & 0.25 \\
 4.72 &  4.14 & 0.22 &  3.84  & 0.26 \\
 5.39 &  3.94 & 0.23 &  3.61  & 0.27 \\
 6.05 &  3.77 & 0.23 &  3.43  & 0.28 \\
 6.71 &  3.63 & 0.23 &  3.27  & 0.28 \\
 7.38 &  3.50 & 0.24 &  3.13  & 0.29 \\
 8.04 &  3.39 & 0.24 &  3.05  & 0.33 \\
 8.71 &  3.29 & 0.25 &  3.05  & 0.43 \\
 9.37 &  3.20 & 0.25 &  3.40  & 0.62 \\
10.03 &  3.12  & 0.25 &  3.44 & 0.65 \\
10.69 &  3.05  & 0.26 &  3.65 & 0.68 \\
11.36 &  2.98  & 0.26 &  3.77 & 0.69 \\
12.02 &  2.92  & 0.26 &  3.77 & 0.70 \\
12.68 &  2.86  & 0.26 &  3.79 & 0.70 \\
13.35 &  2.81  & 0.27 &  3.79 & 0.71 \\
14.01 &  2.76  & 0.27 &  3.79 & 0.72 \\
14.67 &  2.71  & 0.27 &  3.81 & 0.75 \\
15.34 &  2.67  & 0.27 &  3.81 & 0.75 \\
16.00 &  2.63  & 0.27 &  3.81 & 0.75 \\
16.66 &  2.59  & 0.27 &  3.81 & 0.75 \\
\enddata
\end{deluxetable}
\end{document}